\documentclass[usenatbib,usenatbib]{mnras}
\usepackage{todonotes}
\usepackage{times}
\usepackage{amssymb}
\usepackage{amsmath}
\usepackage{graphicx} 
\usepackage{bmpsize}
\usepackage{epstopdf}
\usepackage{dblfloatfix}    
\usepackage{soul}
\usepackage{rotating}
\usepackage{tikz}
\usepackage{soul}

\newcommand{\be}{\begin{equation}}
\newcommand{\ba}{\begin{eqnarray}}
\newcommand{\ee}{\end{equation}}
\newcommand{\ea}{\end{eqnarray}}

\def\gtsima{$\; \buildrel > \over \sim \;$}
\def\ltsima{$\; \buildrel < \over \sim \;$}
\def\gsim{\lower.5ex\hbox{\gtsima}}
\def\lsim{\lower.5ex\hbox{\ltsima}}
\def\simgt{\lower.5ex\hbox{\gtsima}}
\def\simlt{\lower.5ex\hbox{\ltsima}}
\def\simpr{\lower.5ex\hbox{\prosima}}
\def\la{\lsim}
\def\ga{\gsim}

\definecolor{applegreen}{rgb}{0.55, 0.71, 0.0}
\definecolor{ao(english)}{rgb}{0.0, 0.5, 0.0}

\def\simless{\mathbin{\lower 3pt\hbox
   {$\rlap{\raise 5pt\hbox{$\char'074$}}\mathchar''7218$}}}  
\def\simgreat{\mathbin{\lower 3pt\hbox
   {$\rlap{\raise 5pt\hbox{$\char'076$}}\mathchar''7218$}}}  
\def\apj{ApJ}
\def\apjs{ApJS}
\def\apjl{ApJL}
\def\aap{A\&A}

\def\nat{Nature}
\def\mnras{MNRAS}

\title[RSDs in simulations of the 21-cm signal from CD]{Redshift-space distortions in simulations of the 21-cm signal from the cosmic dawn}
\author[H. Ross et al]
{Hannah E. Ross,$^{1}$ \thanks{email: HRoss@lbl.gov} Sambit K. Giri,$^{2}$ Keri L. Dixon,$^{3,4}$ Raghunath Ghara,$^{5,6}$ Ilian T. Iliev,$^{7}$ \\ \\ \LARGE \emph{and Garrelt Mellema $^{8}$}
\\ \\
$^1$ Computational Cosmology Center, Computational Research Division, Lawrence Berkeley National Lab, Berkeley, CA 94720 \\
$^{2}$ Institute for Computational Science, University of Zurich, Winterthurerstrasse 190, 8057
Zurich, Switzerland \\
$^3$ New York University Abu Dhabi, PO Box 129188, Saadiyat Island, Abu Dhabi, United Arab Emirates \\
$^4$ Center for Astro, Particle and Planetary Physics (CAP$^3$), New York University Abu Dhabi \\
$^5$ Department of Natural Sciences, The Open University of Israel, 1 University Road, PO Box 808, Ra'anana 4353701, Israel \\
$^6$ Department of Physics, Technion, Haifa, Israel \\
$^7$ Department of Physics and Astronomy, University of Sussex, Falmer, BN1 9QH, UK \\
$^8$  Department of Astronomy and Oskar Klein Center, Stockholm University, AlbaNova, SE-106 91 Stockholm, Sweden 
}

\date{Accepted ?; Received ??; in original form ???}

\pubyear{2020}

\begin{document}
\label{firstpage}
\pagerange{\pageref{firstpage}--\pageref{lastpage}}
\maketitle

\begin{abstract}

The 21-cm signal from the Cosmic Dawn (CD) is likely to contain large fluctuations, with the most extreme astrophysical models on the verge of being ruled out by observations from radio interferometers. It is therefore vital that we understand not only the astrophysical processes governing this signal, but also other inherent processes impacting the signal itself, and in particular line-of-sight effects. Using our suite of fully numerical radiative transfer simulations, we investigate the impact on the redshifted 21-cm from the CD from one of these processes, namely the redshift-space distortions (RSDs). When RSDs are added, the resulting boost to the power spectra makes the signal more detectable for our models at all redshifts, further strengthening hopes that a power spectra measurement of the CD will be possible. RSDs lead to anisotropy in the signal at the beginning and end of the CD, but not while X-ray heating is underway. The inclusion of RSDs, however, decreases detectability of the non-Gaussianity of fluctuations from inhomogeneous X-ray heating measured by the skewness and kurtosis. On the other hand, mock observations created from all our simulations that include telescope noise corresponding to 1000 h observation with the Square Kilometre Array telescope show that we may be able image the CD for all heating models considered and suggest RSDs dramatically boost fluctuations coming from the inhomogeneous Ly-$\alpha$ background. 

\end{abstract}

\begin{keywords}
cosmology: theory --- radiative transfer --- reionization ---
intergalactic medium --- large-scale structure of universe ---
galaxies: formation 
\end{keywords}

\section{INTRODUCTION}
\label{intro}

The photons emitted from the first luminous sources ended the era known as the Dark Ages, which lasted several million years after the formation of the cosmic microwave background (CMB), and marked the beginning of the Cosmic Dawn (CD). The CD lasted until luminous sources began to significantly ionize the intergalactic medium (IGM), at which point the Epoch of Reionization (EoR) began. Although the IGM had not been significantly ionized during the CD, X-rays from early galaxies were able to penetrate into the IGM and alter its kinetic temperature. Ly-$\alpha$ photons also had long mean free paths and so altered the spin temperature of hydrogen in the IGM. 
Studying the IGM during the CD will shed light on how these first generation sources formed and evolved and how they impacted the IGM of our Universe. See \citet{Pritchard201221cmCentury} and \citet{McQuinn2016TheMedium} for a review on the evolution the IGM during the CD and EoR.

The EoR is constrained to have occurred between $z\sim 5.5$ and 15 by observations of both the CMB \citep[e.g.][]{Komatsu2011Seven-yearInterpretation,PlanckCollaboration2016PlanckHistory} and the Ly-$\alpha$ forest in high redshift quasar spectra \citep[e.g.][]{Fan2005ConstrainingQuasars,Fan2006ObservationalReionization,Davies2018,Bosman2018,Keating2019LongHydrogen}. Hydrogen makes up the vast majority of the IGM, therefore the 21-cm photons originating from the spin-flip transition of neutral hydrogen are expected to be an excellent probe of the EoR and CD. Radio telescopes are currently searching for these redshifted 21-cm photons, an extremely challenging task due to bright astrophysical foregrounds. Current radio telescopes capable of detecting the 21-cm signal include the Giant Metrewave Radio Telescope (GMRT)\footnote{\url{http://www.gmrt.ncra.tifr.res.in/}}\citep{GMRT2013}, the Low Frequency Array (LOFAR)\footnote{\url{http://www.lofar.org/}} \citep{vanHaarlem2013LOFAR:ARray}, the Murchison Widefield Array (MWA)\footnote{\url{http://www.mwatelescope.org/}} \citep{bowman13} and the Hydrogen Epoch of Reionization Array (HERA)\footnote{\url{https://reionization.org/}} \citep{2017PASP..129d5001D}. These radio interfermometers are sensitive to the fluctuations in the signal rather than its magnitude. To date, there has not yet been a detection of these fluctuations, although some upper limits have been established by LOFAR \citep[e.g.][]{2019ApJ...883..133K,Mertens2020ImprovedLOFAR,Gehlot2020} and MWA \citep[e.g.][]{Barry2019,Trott2020DeepObservations} that have lead to some constraints on the signal \citep[e.g.][]{Ghara2020ConstrainingObservations,Mondal2020TightLOFAR,2020arXiv200603203G,Greig2020ExploringSignal}. The Square Kilometre Array (SKA)\footnote{\url{https://www.skatelescope.org/}}\citep{Mellema2013ReionizationArray,Koopmans2015TheArray} radio telescope will become operational during the second half of the current decade. The low frequency component of SKA (SKA-Low) will be sensitive enough to not only detect the power spectrum of the 21-cm signal during CD/EoR but also produce images. 

Single antenna telescopes are searching for the global 21-cm signal, such as the Experiment to Detect the Global EoR Signature (EDGES)\footnote{\url{https://www.haystack.mit.edu/ast/arrays/Edges/}} \citep{Bowman2018AnSpectrum}, the Large-aperture Experiment to Detect the Dark Age (LEDA)\footnote{\url{http://www.tauceti.caltech.edu/leda/}} \citep{price2018design}, Shaped Antenna measurement of the background RAdio Spectrum (SARAS2)\footnote{\url{http://www.rri.res.in/DISTORTION/saras.html}} \citep{Singh2018}  and the 
Dark Ages Polarimeter Pathfinder (DAPPER)\footnote{\url{https://www.colorado.edu/ness/dark-ages-polarimeter-pathfinder-dapper}} \citep{burns2019}.  \citet{Bowman2018AnSpectrum} claim to have detected a 21-cm absorption signal in the EDGES low-band antenna at $z\sim 17$. The magnitude of this signal is far greater than expected with our current theoretical understanding, therefore the accuracy of these results has been debated \citep[e.g.\ in][]{Hills2018ConcernsData,2018ApJ...858L..10D,2019ApJ...880...26S,2019ApJ...874..153B}. If correct, these results suggest our understanding of the early Universe needs to be updated to include either a strong radio background other than the CMB \citep[][]{2018ApJ...858L..17F, 2019MNRAS.486.1763F} or an additional cooling mechanism \citep[see e.g.,][]{2014PhRvD..90h3522T,Bowman2018AnSpectrum,2018PhRvL.121a1101F,2018arXiv180210094M}. 

SKA-Low will observe the sky at various frequencies, which will correspond to different redshifts. A sequence of images at different redshifts will form three-dimensional data, known as the tomographic data set \citep[we refer the interested reader to][]{giri2019tomographic}. Hence, when analysing the redshifted 21-cm signal, it is important to take into account line-of-sight (LoS) effects that alter the signal independent of astrophysical processes. The redshifted 21-cm signal will be impacted by two LoS effects: redshift-space distortions \citep{Jensen2013ProbingDistortions,Mao2012Redshift-spaceRe-examined}, and the light-cone effect \citep{Datta2012Light-coneSpectrum}. Our imperfect knowledge of the underlying Cosmology of the Universe can add the Alcock-Paczynski effect to the redshifted 21-cm signal \citep{Alcock1979,Nusser2005TheMaps}. The scope of this work is limited to RSDs and we leave a study of the light-cone and Alcock-Paczynski effects for future work.

In our expanding Universe, cosmological redshift corresponds to a certain distance. Peculiar velocities of the hydrogen gas can add additional blue or redshifts to the signal which then change the apparent distance to the observed medium, thus distorting the the distribution of the signal in the tomographic data set. This effect is called the redshift-space distortion (RSD) \citep[e.g.][]{Jensen2013ProbingDistortions,Majumdar2015EffectsDistortions}. 

The key difference between the 21-cm signal in real-space and redshift-space caused by RSDs is the \textit{Kaiser effect} \citep{Kaiser1987}, in which dense regions appear denser and under-dense regions appear more diffuse than they are in reality. The Kaiser effect is caused by the tendency for matter to move towards high density regions. Radiation emitted from gas between us and a dense area will generally be red-shifted as the gas is falling away from us into the high-density region and radiation emitted from gas on the far side will be blue-shifted as the gas is falling towards us. While the density fluctuations dominate (i.e. in the absence of an in-homogeneous Ly-$\alpha$ background, X-ray heating and ionization) the Kaiser effect boosts fluctuations in the 21-cm power spectrum on all scales by a factor of 1.87 \citep{Bharadwaj2004TheReionization,Barkana2005AFluctuations}.

Previous studies have shown that the RSD can amplify the 21-cm power spectrum up to a factor of $\sim 5$ \citep{Mao2012Redshift-spaceRe-examined,Ghara201521cmVelocities}. Including RSDs in the simulated 21-cm signal is important to generate realistic results to compare to observations \citep{Jensen2013ProbingDistortions,Jensen2016TheMeasurements}. Most previous work has focused on the impact of RSDs of the 21-cm signal from the EoR \citep{McQuinn2006CosmologicalReionization,Mesinger2011,Jensen2013ProbingDistortions,Jensen2016TheMeasurements}, with only a few previous works including RSDs in the CD \citep[e.g][]{Fialkov2014,Ghara201521Effects}. Other work has focused specifically on using the anisotropy generated in the 21-cm signal by the RSDs to separate astrophysical and cosmological affects \citep[e.g.][]{Shaw2008,Shapiro2013,Majumdar2015EffectsDistortions}. 

In this paper we will focus on the RSDs in the CD. Previous works that include RSDs in this epoch have been limited to semi-numerical simulations. In this paper, for the first time, we will build on these studies by looking at the impact of including the RSDs on the 21-cm signal from the CD with fully numerical simulations. As well as repeating the analysis done for these previous simulations (the isotropic 21-cm power spectra, one-point statistics and images), this time including RSDs, we will also investigate the anisotropy of the 21-cm signal. We will also investigate the detectability of the 21-cm signal using SKA (as was done for the detectability of the EoR for LOFAR in \citet{Jensen2013ProbingDistortions}). Finally we will investigate the impact of the RSDs on a variety of astrophysical scenarios in the CD. 

The structure of the paper is as follows. In Section~\ref{sec:method}, we outline the $N$-body and radiative transfer (RT) simulations used in this work and the source models they employ. In Section~\ref{sec:method}, we summarize the extraction of the 21-cm signatures and relevant statistics from our simulations and detail our calculations of the LoS effects. Section~\ref{sec:results} contains our results, primarily comparisons between the different source models. We then summarize our findings in Section~\ref{sec:conclusions}. The cosmological parameters we use throughout this work are ($\Omega_\Lambda$, $\Omega_\mathrm{M}$, $\Omega_\mathrm{b}$, $n$, $\sigma_\mathrm{8}$, $h$) = (0.73, 0.27, 0.044, 0.96, 0.8, 0.7); where the notation has the usual meaning and $h = \mathrm{H_0} / (100 \  \mathrm{km} \ \rm{s}^{-1} \ \mathrm{Mpc}^{-1}) $. These values are consistent with the final results from WMAP \citep{Komatsu2011Seven-yearInterpretation} and Plank combined with all other available constraints \citep{PlanckCollaboration2016PlanckHistory,PlanckCollaboration2018PlanckParameters}.

\section{Methodology}
\label{sec:method}

\subsection{Simulations}
\label{sec:sims}

A high-resolution, $N$-body simulation run with \textsc{\small CubeP$^3$M} Code \citep{Harnois2013} is used to give the density fields and halo catalogues. This simulation has a  $349\,$~co-moving~Mpc per side volume to capture large-scale fluctuations with $4000^3$ particles to enable reliable halo identification down to $10^9\, \rm M_\odot$ with the halo finder described in \citet{Watson2013TheAges}. Haloes  down to $10^8\, \rm M_\odot$ were added using a subgrid model from \citet{Ahn2015a}. The $N$-body particles are smoothed into a $250^3$ grid for the RT using SPH-like kernel \citep[][]{Iliev2014SimulatingEnough}. For more details on this simulation see \citet{Dixon2016TheReionization}. 

The RT simulations were run using the \textsc{\small C$^2$-Ray} code \citep{Mellema2006CRadiation,Friedrich2012} over a succession of density grids and source lists derived from the $N$-body simulation above and updated every 11.52~Myrs. These simulations include multi-frequency heating, all three species of helium, secondary ionizations, and a multi-phase approach, described in \citet{Ross2019EvaluatingDawn}. A total of five simulations are analysed, although most of the paper focuses on the original simulation presented in \citet{Ross2017}. The details of the source models are outlined in Section~\ref{sec:sources}. These simulations have previously been presented in \citet{Ross2017}, \citet{Ross2018} and \citet{Ross2019EvaluatingDawn}.

To model the inhomogeneous Ly-$\alpha$ flux background we use a semi-numerical method as a post-processing step, which is outlined in \citet{Ghara2016} and \citet{Ross2018}. This method assumes a spherical profile (showed to be a good approximation even in the presence of density fluctuations by \citet{Vonlanthen2011} and drops off as $1/r^2$ \citep{semelin07,Vonlanthen2011,Higgins2012}.

\subsection{Sources}
\label{sec:sources}

\begin{table*} 
\begin{center}
\caption {The astrophysical parameters of our different models including the spectral indecies of our X-ray sources in the different and Ly-$\alpha$ models described in Section~\ref{sec:sources}. }
\begin{tabular}{ | l | c | c | c | c | c | c | c | c | c | r |}
\hline
  & \  \textbf{S1 EL} \  & \  S1 LL \  & \  S2 EL \  & \  S2 LL \  & \  S3 EL \  & \  S3 LL \  & \  S4 EL \  & \  S4 LL \\
\hline \hline
\ \ \ \ \ \ \ \ \ $\alpha^{\rm h}_{\rm x}$ & \textbf{1.5} & 1.5 & 1.5 & 1.5 & -- & -- & -- & -- \\ 
\hline
\ \ \ \ \ \ \ \ \ $\alpha^{\rm q}_{\rm x}$ & \textbf{--} & -- & 0.8 & 0.8 & 0.8 & 0.8 & 1.6 & 1.6 \\ 
\hline
Ly-$\alpha$ &\textbf{Early} & Late & Early & Late & Early & Late & Early & Late \\
\hline
\label{tab:models}
\end{tabular}
\end{center}
\end{table*}

We consider three categories of luminous sources: stellar sources, high-mass X-ray binaries (HMXBs), and quasars (QSOs). The models used are summarised in Table~\ref{tab:models} with the fiducial model emboldened. The majority of this work (Section~\ref{sec:rsd}) focuses on S1~EL (emboldened in Table~\ref{tab:models}), while the other models are analysed in Section~\ref{sec:allmodels} to determine their differences and detect-ability.

\subsubsection{Stellar sources}

Stellar sources form inside the dark matter haloes obtained from the $N$-body simulation. High-Mass Atomically Cooling haloes, HMACHs, are haloes with masses large enough ($M>10^9$~M$_\odot$) to atomically cool ionized gas and so are never suppressed. Haloes of lower mass are not able to sufficiently cool ionized gas from the IGM, however, they can accrete neutral gas.  We assume that these Low-Mass Atomically Cooling haloes, LMACHs, do not form stars when located in cells that are more than 10 per~cent ionized.  Our stellar sources have black-body spectra with an effective temperature of $T_{\mathrm{eff}} = 5 \times 10^4$~K. For further details on the implementation of these sources see \citet{Dixon2016TheReionization}.

Stellar sources also produce Ly-$\alpha$ photons, which impact the spin temperature (see Section~\ref{sec:dTb}). We consider two scenarios for the timing of Ly-$\alpha$ saturation: \newline 
\textbf{1) Early:} where the Ly-$\alpha$ background is already built up by very early sources called minihaloes before the simulation starts. Here we do not need to calculate the Lyman-$\alpha$ flux. \newline 
\textbf{2) Late:} where the only Ly-$\alpha$ emitters are the sources included in our simulations. Here we use the semi-numerical method described above at the end of Section~\ref{sec:sims}. The spectra for the Ly-$\alpha$ flux calculations were obtained from the stellar population synthesis code \textsc{\small PEGASE2} \citep{Fioc1999}. Stars are assumed to be between 1 and 100~M$_\odot$ with a Salpeter initial mass function, with low metalicities (1000 times less than that of the sun). For these calculations we assume all Lyman-$\alpha$ photons escape, and 10~per~cent of ionizing photons.

\subsubsection{HMXBs}

The HMXB sources are also assumed to form in dark matter haloes and to have power-law spectra:
\begin{equation}
L_{\rm h}(\nu) \propto \nu^{-\alpha^{\rm h}_{\rm x}},
\end{equation}
where $\alpha^{\rm h}_{\rm x} = 1.5$ (in agreement with \citet{Hickox2007}). The energy range extends from 272.08 eV, which is in agreement with observational works on the obscuration \citep[e.g.][]{Lutivnov2005} and gamma-ray bursts  \citep[e.g.][]{Totani2006,Greiner2009}, to 100 times the second ionization of helium (5441.60 eV). As with the stellar sources, the luminosity is related to the mass of the host halo, with a lower photon efficiency and are roughly consistent with measurements of X-ray binaries in local, star-bursting galaxies \citep{Mineo2012}. More details of these sources are available in \citet{Ross2017}.
 
\subsubsection{QSO sources}

QSOs also have a power-law spectrum, given by:
\begin{equation}
L_{\rm q}(\nu) \propto \nu^{-\alpha^{\rm q}_{\rm x}},
\end{equation} 
where $\alpha^{\rm q}_{\rm x}$ = 0.8 \citep{Ueda2014} or 1.6 \citep{Brightman2013} for our two QSO models. Other than having a power-law spectra, our QSO sources behave very differently from HMXBs and stellar sources. We assume that our QSO-like sources only produce X-rays and the X-ray emissivity from QSOs is quantified using the QXLF from \citet{Ueda2014}. The number density of QSOs in our simulation volume, $n_{\rm q}$, is calculated by integrating the QXLF, $\Phi(L,z)$, for each redshift and then randomly sampling the QXLF to assign luminosities. Note that this approach produces many more QSOs than \citet{Ueda2014} would predict and has been calculated this way as the goal of running these simulations was to consider the maximum impact QSOs could have. There is a rapid drop in QSO number density at high redshifts driven by the lower HMACH density. These QSOs are then randomly placed in HMACHs and remain active for three time steps (34.56~Myr), with the luminosity being reassigned every time step. Many more details on this implementation are available in \citet{Ross2019EvaluatingDawn}.

\subsection{Differential brightness temperature}
\label{sec:dTb}

In this section, we give a brief theory of 21-cm signal. The radio telescopes record a quantity known as the differential brightness temperature. Section~\ref{sec:dTb} will describe the differential brightness temperature. In the following section, we will present the method used to add RSDs from to simulation outputs.

The 21-cm differential brightness temperature is observed against a radio background, which in this case is the CMB radiation. The differential brightness temperature can be given as \citep[see e.g,][]{Madau199721Redshift,Furlanetto2006CosmologyUniverse}, 
\begin{eqnarray}
 \delta T_\mathrm{b} (\mathbf{x}, z) \!  & \approx & \! 27 ~ x_{\rm HI} (\mathbf{x}, z) [1+\delta_{\rm B}(\mathbf{x}, z)] \left(\frac{\Omega_\mathrm{B} h^2}{0.023}\right) \nonumber\\
&\times& \!\left(\frac{0.15}{\Omega_\mathrm{m} h^2}\frac{1+z}{10}\right)^{1/2}  \left(1-\frac{T_\mathrm{CMB}(z)}{T_\mathrm{S}(\mathbf{x}, z)} \right)\,\rm{mK},
\nonumber \\
\label{eq:brightnessT}
\end{eqnarray}
where the quantities  $x_{\rm HI}$, $\delta_{\rm B}$ and $T_\mathrm{CMB}(z) = 2.725~(1+z)$ K denote the neutral hydrogen fraction, baryonic density contrast and the CMB temperature, respectively, each at position $\mathbf{x}$ and redshift $z$. $T_\mathrm{S}$ represents the spin temperature of hydrogen in the IGM.

The $T_{\rm S}$ reflects the relative number of atoms in the singlet and triplet state of the 21-cm line, given by \citep{Field1958,Pritchard201221cmCentury}:
\begin{equation}
T_{\rm S} = \frac{T_{\rm CMB}+y_\alpha T_\alpha + y_{\rm c} T_{\rm K}}{1+y_\alpha+y_{\rm c}},
\end{equation}
where $T_\alpha$ is the radiation colour temperature, $T_{\rm K}$ is the kinetic temperature of the gas, and $y_\alpha$ and $y_{\rm c}$ are the coupling coefficients corresponding to the Ly-$\alpha$ decoupling (the Wouthuysen-Field effect) and collisional decoupling, respectively.  

We do not include the collisional coupling as the Universe has expanded enough that the density of the IGM is insufficient to produce non-negligible collisions between hydrogen atoms, and we do not resolve the dark matter filaments with sufficient densities for collisional coupling to become efficient on the RT grid. The $y_\alpha$ is calculated using:
\begin{equation}
y_\alpha = \frac{T_*}{T_\mathrm{K}} \frac{P_{10}}{A_{10}},
\end{equation}
where $P_{10}$ is the radiative de-excitation rate due to Ly-$\alpha$ photons ($\sim 10^9 J_\alpha$, where $J_\alpha$ is the Ly-$\alpha$ flux), and $T_*=0.068$~K is the temperature corresponding to the energy difference between two the states.

\subsection{Redshift-space distortions}
\label{sec:los-effects}
A 3D data set of the 21-cm signal with radio observations is formed by observing the signal at different frequencies. The direction of the data set along which observation frequency evolves is the LoS direction.
Therefore we assume one axis of our simulated 21-cm signal data set to correspond to the LoS direction.
Each slice along the LoS direction will correspond to the signal at a redshift $z$. However, due to RSD the signal at $z$ will be observed at the following frequency,
\begin{eqnarray}
\nu_\mathrm{obs} = \frac{\nu_\mathrm{21}(1-v_\parallel/c)}{1+z} \ ,
\end{eqnarray}
where $\nu_\mathrm{21}$, $v_\parallel$ and $c$ are the rest-frame frequency of the 21-cm signal, peculiar velocity and speed of light respectively. The impact of RSD is added to the $\delta T_\mathrm{b}$ coeval cubes using the MM-RRM scheme presented in \citet{Mao2012Redshift-spaceRe-examined}. We follow the implementation of the scheme provided in \citet{Jensen2013ProbingDistortions} and \citet{Datta2014LightImplications}. The modules to implement RSD to them are provided in a publicly available package called {\sc Tools21cm} \citep{giri2020tools21cm}. Recently, \citet{chapman2019AFullcone} presented a new method to implement RSD in 21-cm light cones. They do not carry out a quantitative comparison of the different methods but qualitatively, the impact of RSDs the signal in this work agrees with their results. The reader should note that we add RSD in coeval cubes as the signal in light cone will be prone to the light cone effect \citep[][]{Ghara201521Effects}, which is beyond the scope of this paper.

\subsection{Telescope Noise}
\label{sec:noise}
\begin{table}
	\centering
	\caption{The parameters used in this study to model the telescope properties.}
	\label{tab:telescope_param}
	\begin{tabular}{lccccc}
		\hline
		Parameters & Values \\
		\hline
		\hline
		Maximum baseline ($B$) & 1.2 km  \\
        Observation time ($t_\mathrm{int}$)& 1000 h  \\
        System temperature ($T_\mathrm{sys}$) & $60 (\frac{\nu}{300\mathrm{MHz}})^{-2.55}$ K  \\
		Effective collecting area ($A_\mathrm{D}$) & 962 $\mathrm{m}^2$  \\
        Declination & -30$^\circ$ \\
        Observation hour per day & 6 hours \\
        Signal integration time & 10 seconds \\
		\hline
	\end{tabular}
\end{table}

We follow the method given in \citet{Ghara2017ImagingSKA}  and \citet{Giri2018OptimalObservations} to simulate the expected noise from the first generation of SKA-Low (hereafter SKA1-Low).  Even though the construction of SKA1-Low has not started, the distribution of antennae is being planned.\footnote{The most recent antennae configuration can be found at \url{https://astronomers.skatelescope.org/documents/}.} The parameters values describing the properties of telescope are given in Table~\ref{tab:telescope_param}. These telescope parameters are similar to the ones used in \citet{Giri2019}, which explores the end of reionization. In this study, we look at earlier redshifts and therefore the noise level will be higher than the noise level during EoR. 

Our telescope noise simulation gives us noise maps at each redshift for 1000 h. At $z\sim 15$, the standard deviation of the simulated telescope noise ($\sigma_\mathrm{noise}$) at the resolution of simulation ($\Delta x_\mathrm{sim}\sim 1.4$~Mpc) is $\sim 29~ \mathrm{K}$, which is much higher than the standard deviation of our signal. In order to study the expected images produced by SKA1-Low, we have to reduce the resolution. We decrease the resolution of the images to the scale corresponding to a maximum baseline of 1.2 km, which is $\Delta x_\mathrm{SKA}\sim 29.8$ Mpc. The $\sigma_\mathrm{noise}$ in the degraded resolution is $\sim 3~ \mathrm{mK}$. We refer the reader to figure~2 in \citet{Mellema2015HISKA} to see how $\sigma_\mathrm{noise}$ depends on the resolution of 21-cm images observed with SKA1-Low.

\section{Results}
\label{sec:results}

\subsection{Impact of redshift-space distortions on the fiducial model}
\label{sec:rsd}

We first discuss the impact of including RSDs on the results of our fiducial simulation (shown in bold in Table~\ref{tab:models}) for clarity. This simulation only contains HMXBs, which results in reasonably early heating (temperature saturation is reached at around $z\sim12.5$ and the transition from absorption to emission for the average value of the signal happens at around $z\sim14.0$. In this model, it is assumed that Ly-$\alpha$ saturation has already been reached before the simulation has started.

\begin{figure}
  \centering
  \begin{tabular}{c}
    \includegraphics[width=1.\columnwidth]{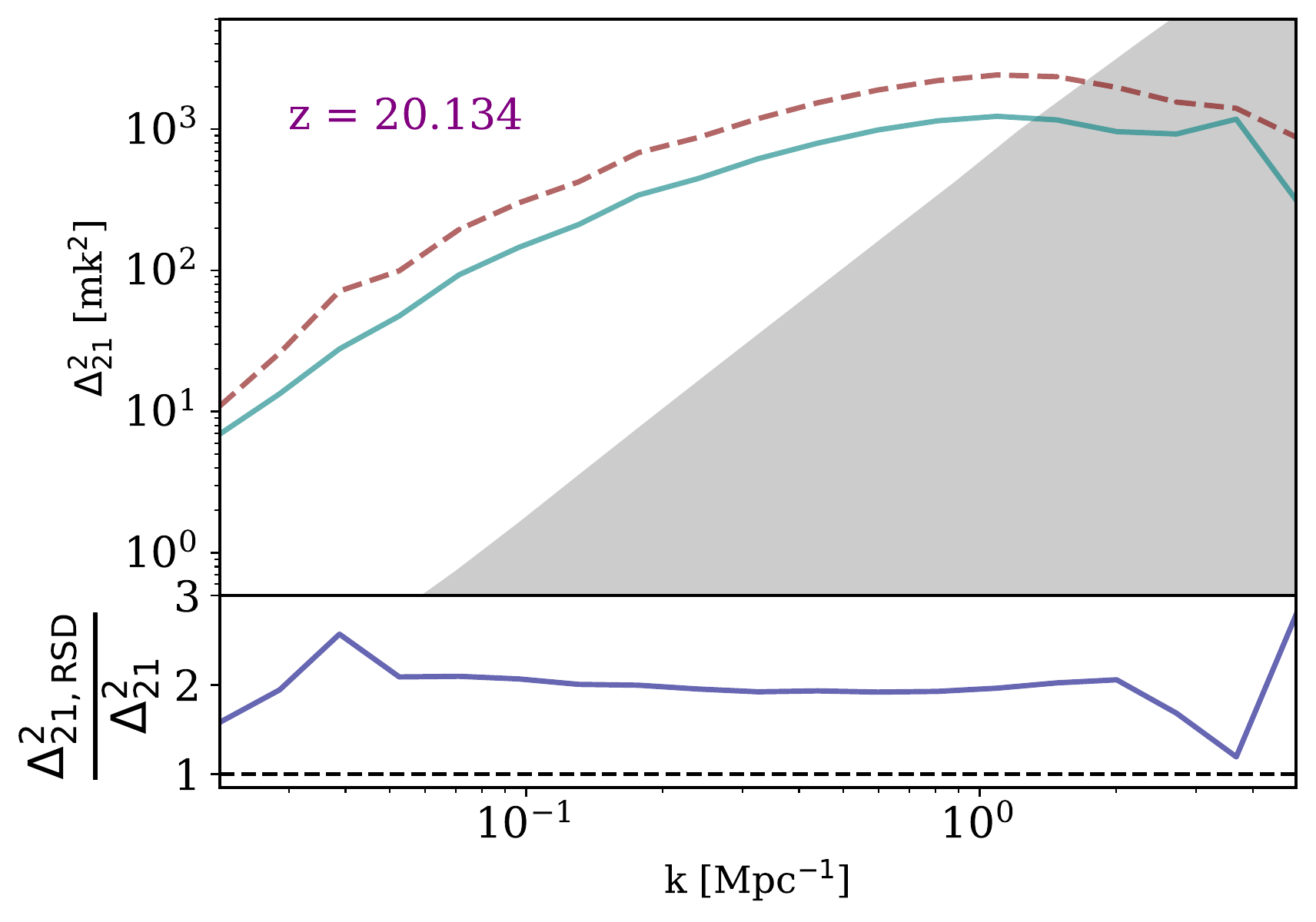} \\

     \includegraphics[width=1.\columnwidth]{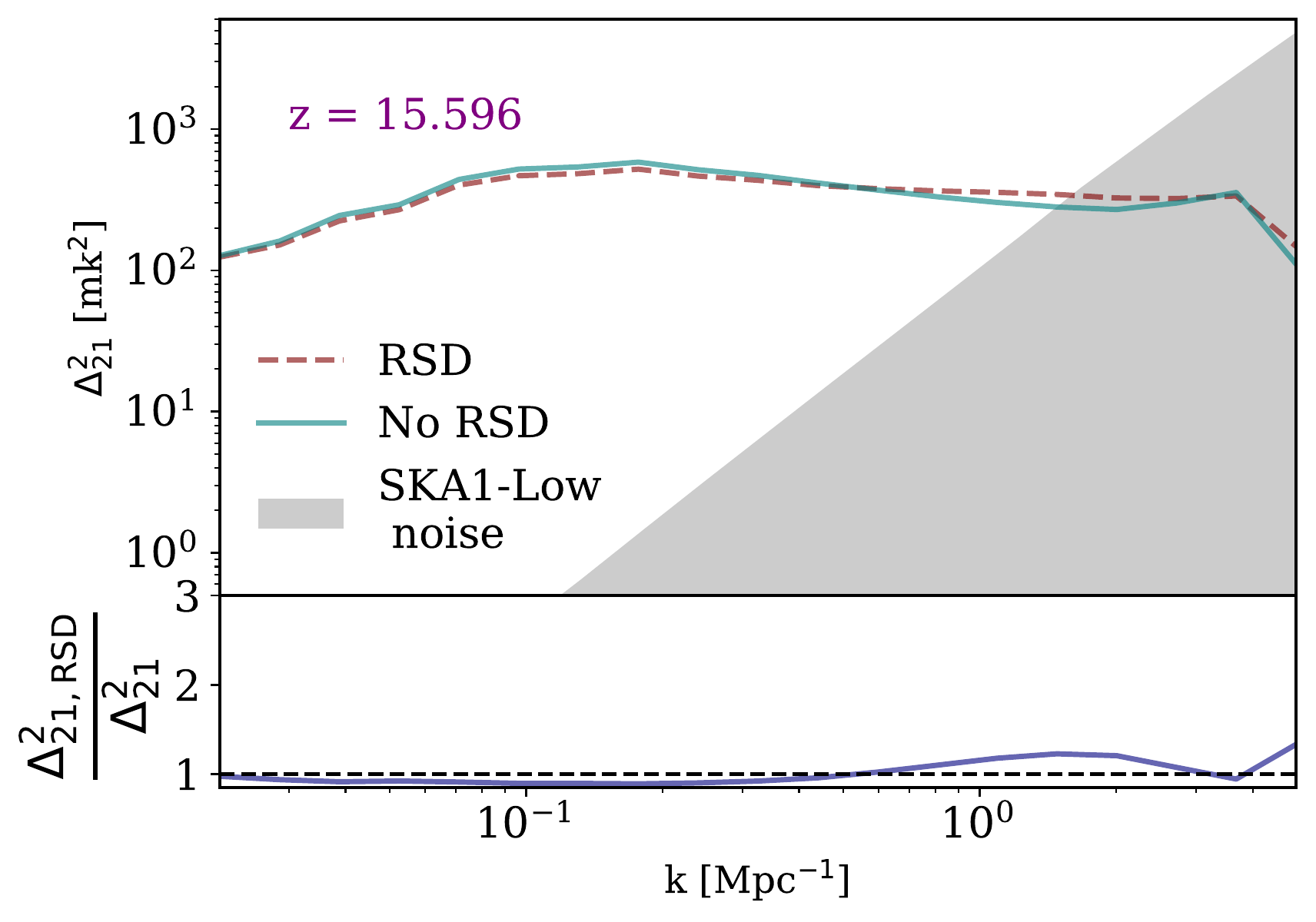} \\

      \includegraphics[width=1.\columnwidth]{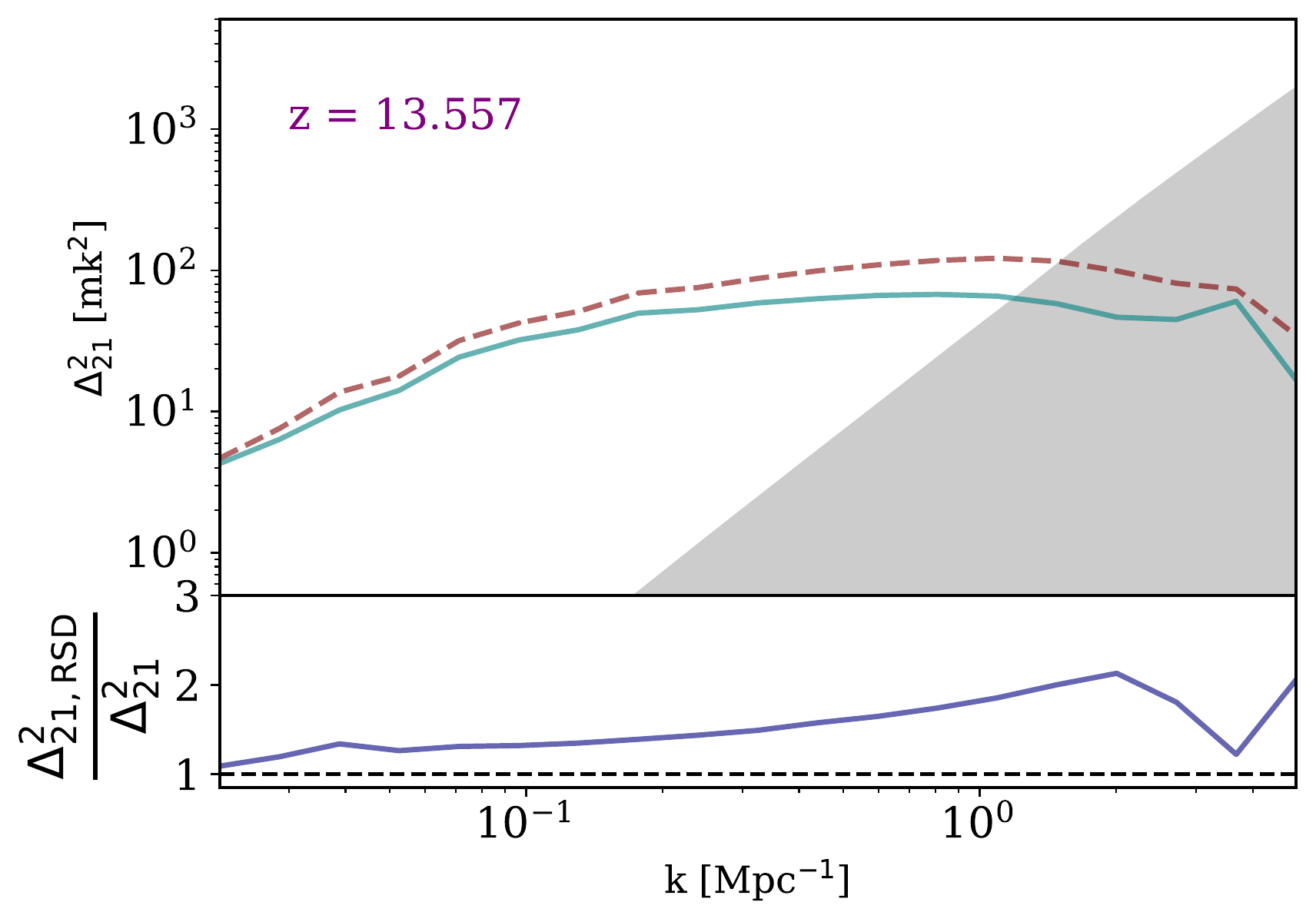} \\
  \end{tabular} \\
  \caption{The impact on $\Delta^2_\mathrm{21}$ when RSDs are included for a selection of redshifts with each plot being split into two panels. Upper panels: $\Delta^2_\mathrm{21}$ with (dashed red line ) and without (solid teal line) RSDs. The noise dominated region from SKA1-low for 1000h (See Table~\ref{tab:telescope_param} for other parameters) is represented by the shaded region. Lower panels: the ratio of $\Delta^2_\mathrm{21}$ with and without RSDs. RSDs boost the power on all scales before significant in-homogeneous X-ray heating begins and once temperature saturation is approached. However, this boost vanishes at intermediate times as the effects of X-ray heating and the kaiser interfere destructively.}
  
  \label{fig:ps_rsd}
\end{figure}
\begin{figure}
  \centering
  \begin{tabular}{c}
    \includegraphics[width=1.\columnwidth]{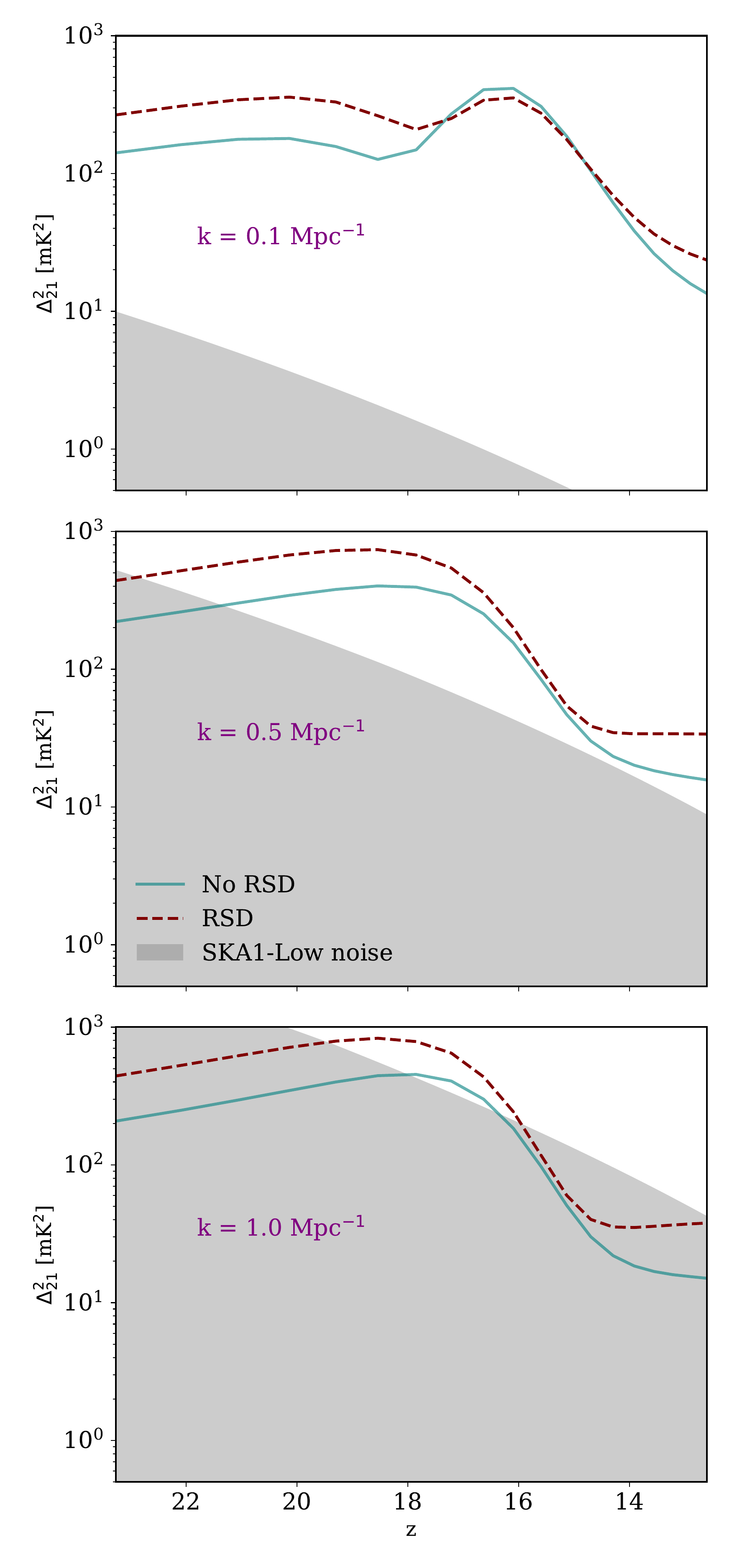}
  \end{tabular}
  \caption{The evolution of three separate wave-numbers (0.1~Mpc$^{-1}$, 0.5~Mpc$^{-1}$, and 1.0~Mpc$^{-1}$ in the top, middle, and lower panels respectively) over time with (dashed red line) and without (solid teal line) RSDs. The noise dominated region from our noise model (See Table~\ref{tab:telescope_param} for parameters) is shaded in grey. RSDs boost the power on all scales, except when the fluctuations in the signal are dominated by fluctuations in the kinetic temperature at which point the fluctuations from X-ray heating and the kaiser effect cancel each other out. During this time increase in power becomes less prominent on small scales and vanishes completely on large scales.}
  
  \label{fig:kmodes_rsd}
\end{figure}

\subsubsection{The spherically averaged power spectra}
 
The primary goal of the current radio telescopes is to measure the fluctuations in $\delta T_\mathrm{b}$. The first-order metric to quantify the fluctuations is the power spectrum. Assuming the signal to be isotropic, we can estimate the spherically averaged power spectrum, which is defined as
\begin{eqnarray}
\langle \hat{\delta T_\mathrm{b}}(\mathbf{k}) \hat{\delta T_\mathrm{b}}^*(\mathbf{k'})\rangle = 2\pi \delta_\mathrm{D}(\mathbf{k}-\mathbf{k'}) P(k),
\end{eqnarray}
where $\hat{\delta T_\mathrm{b}}(\mathbf{k})$ is the Fourier transform of the $\delta T_\mathrm{b} (\mathbf{r})$ and $\delta_\mathrm{D}$ is the three dimensional Kronecker delta. In this paper, we work with dimensionless power spectrum $\Delta^2_\mathrm{21}(k) = P(k)k^3/(2\pi^2)$ \citep{Peacock_1999}. Hereafter, we refer $\Delta^2_\mathrm{21}$ as the power spectrum unless specified.

In Figure~\ref{fig:ps_rsd}, the $\Delta^2_\mathrm{21}$ with (dashed red line) and without (solid blue line) RSDs are shown in the upper half of each panel along with the expected noise from SKA1-Low (shaded region) for a selection of insightful redshifts. The lower half of the panels display the ratio of $\Delta^2_\mathrm{21}$ with and without RSDs. RSDs boost the signal's power spectrum significantly before and after the transition from emission to absorption. This boost in $\Delta^2_\mathrm{21}$ is due the Kaiser effect, causing an increase in fluctuations and therefore in power.  The ratio plot for $z =20.134$ (before heating has started) and $z=13.557$ (where temperature saturation has been reached) approximately matches the amplification factor of 1.87, calculated by \citet{Mao2012Redshift-spaceRe-examined}. This match is expected as the fluctuations $\delta T_{\mathrm{b}}$ are largely unaffected by temperature variations, arising mainly from fluctuations in the density field. At $z=15.596$, X-ray heating causes the signal coming from high-density peaks to increase as the temperature goes up, effectively counteracting the Kaiser effect. On very large scales, the inclusion of RSDs causes a small decrease in power, likely due to these regions continuing to appear more void-like than they are in real-space. The Kaiser effect returns by the end of the simulation, once temperature saturation has been reached as the dense regions again have an impact on the signal.

The impact of RSDs can be seen clearly in the evolution of individual wave-numbers shown in Figure~\ref{fig:kmodes_rsd}.
Although the additional fluctuations from RSDs marginally decrease the power on the very large scales during inhomogeneous heating ($k = 0.1~\mathrm{Mpc}^{-1}$ in the top panel, $z\sim17.5$ to $15.5$), the fluctuations are boosted for other redshifts. The small decrease in power on large-scale fluctuations during X-ray heating is unlikely to significantly impact the detect-ability for the SKA1-Low telescope noise and, for the noise scenario we consider here, these high wave-numbers are well above the expected noise. The smaller scales shown here ($k=0.5~\mathrm{Mpc}^{-1}$ in the middle panel and $k=1.0~\mathrm{Mpc}^{-1}$ in the lower panel) $\Delta^2_\mathrm{21}$ is boosted for all redshifts by RSDs, increasing their detectability. When RSDs are included, $\Delta^2_\mathrm{21}$ for the intermediate scale ($k=0.5~\mathrm{Mpc}^{-1}$) is above the telescope noise for all $z\lesssim 22.5$ for the noise model considered here. However, even in this optimistic noise scenario $\Delta^2_\mathrm{21}$ is only above telescope noise for $z\sim19-16$ for the smallest scale shown here ($k= 1.0~\mathrm{Mpc}^{-1}$).

\begin{figure} 
    \includegraphics[width=1.\columnwidth]{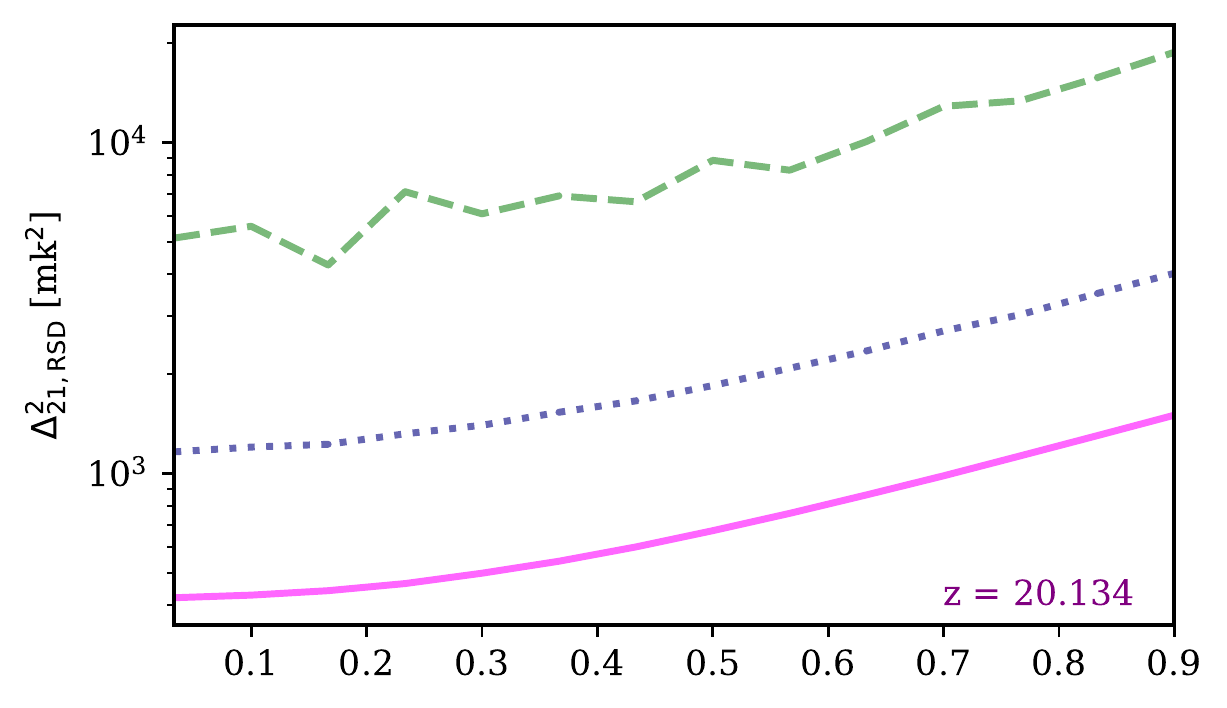} \\
    \includegraphics[width=1.\columnwidth]{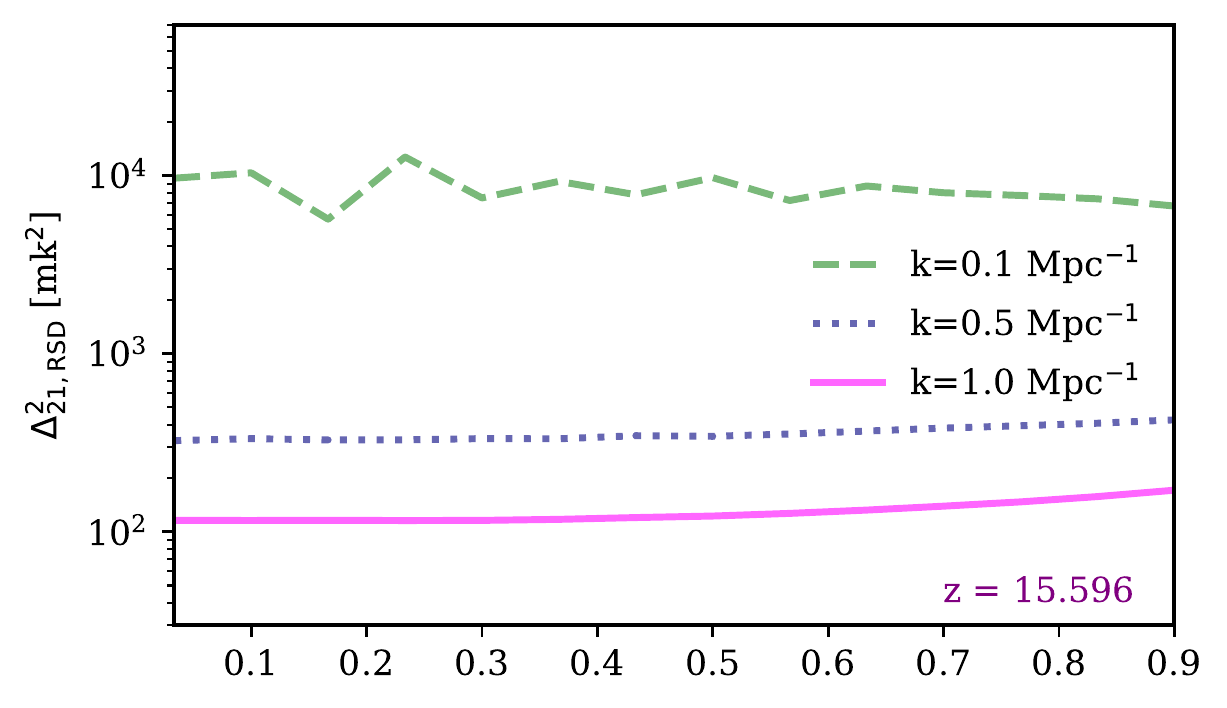} \\
    \includegraphics[width=1.\columnwidth]{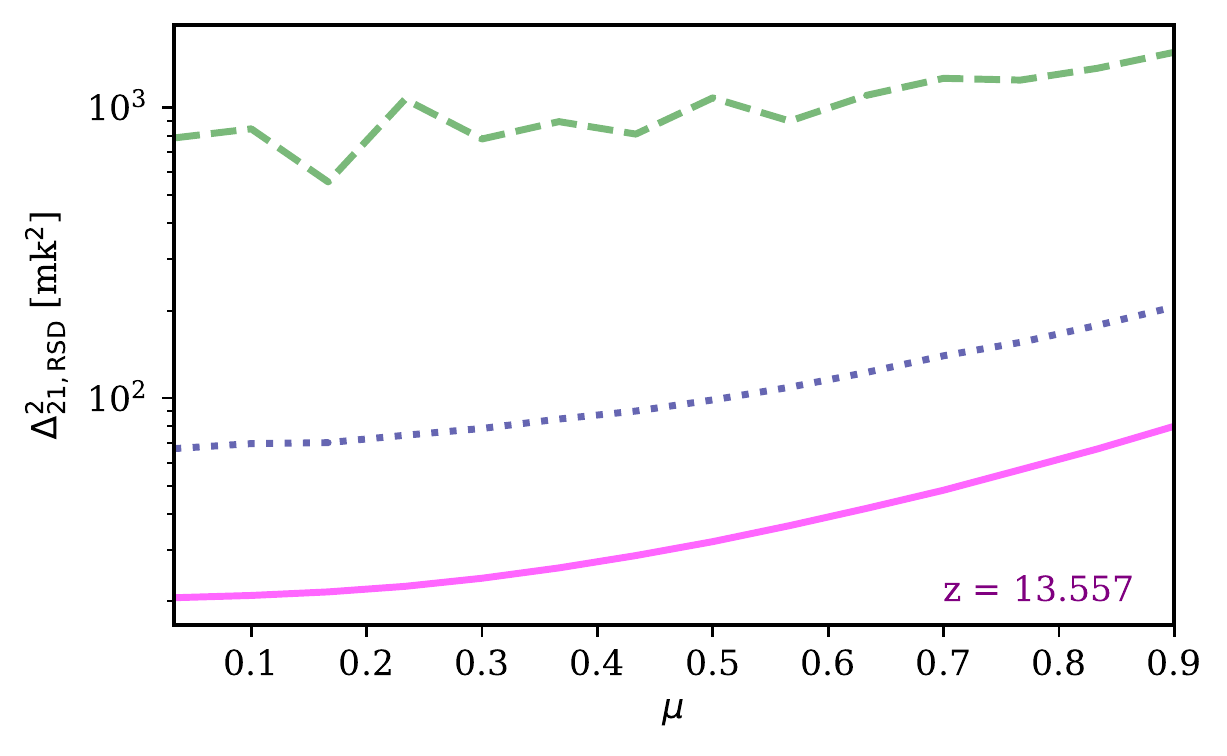}
  \caption{The power spectrum dependence on $\mu$ for a selection of different redshifts and wave-numbers (0.1~Mpc$^{-1}$, 0.5~Mpc$^{-1}$ and 1.0~Mpc$^{-1}$ are represented by dashed green lines, dotted navy lines and solid magenta lines respectively) for simulations including RSDs. Note that for simulations without RSDs the signal is isotropic so $\Delta^2_\mathrm{21}$ would not vary with $\mu$. There is a clear $\mu$ dependence before X-ray heating gets underway and once temperature saturation is approached. However, this dependence vanishes while the fluctuations are dominated by X-ray heating. This lack of anisotropy is due to the fluctuations from X-ray heating counteracting additional fluctuations from RSDs.}
  \label{fig:mu_k_ps}
\end{figure}
 
\begin{figure}
  \centering
    \includegraphics[width=0.99\columnwidth]{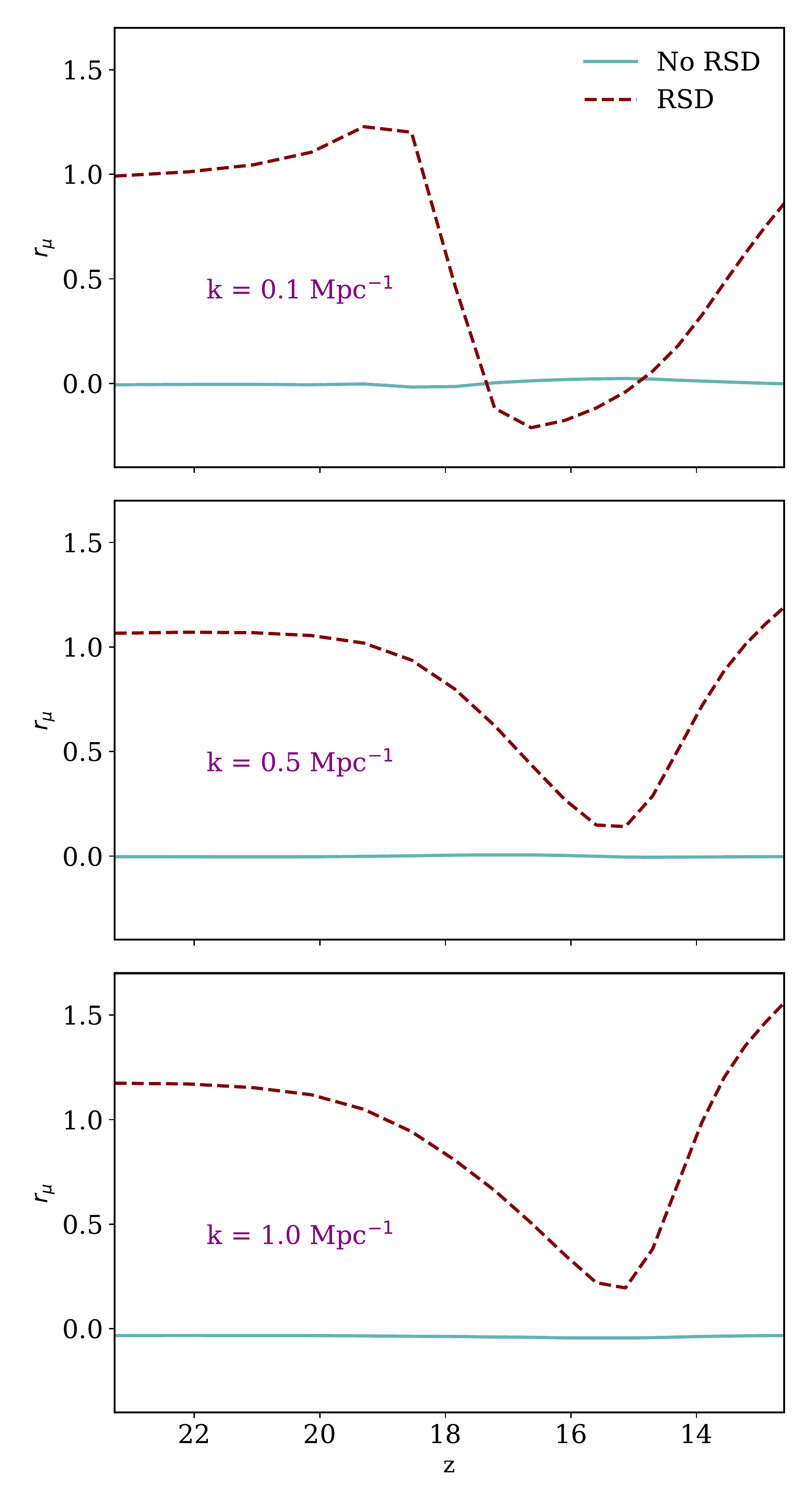}\\
  \caption{The evolution of anisotropy ratio ($\mathrm{r}_\mu (k,z)$, see \citet{Fialkov2015}) over time to see the impact of RSDs on the anisotropy during the CD. As expected, $\mathrm{r}_\mu (k,z)$ is zero when RSDs are not included (solid teal lines) because the signal is isotropic. When RSDs are included (dashed red lines) the anistropy ratio shows a trough for all scales (occurring slightly sooner for the largest scale shown here, 0.1~Mpc$^-1$ than the other scales) when X-ray heating is underway (but temperature saturation has not been reached). This trough demonstrates how the anistropy in the signal is removed by the fluctuations from X-ray heating effectively cancelling the fluctuations from RSDs.}
  \label{fig:anistropy_ratio}
  \end{figure}

\subsubsection{Anisotropy in the power spectrum}

The spherically averaged value of $\Delta^2_\mathrm{21}$ accurately measures the power spectra in real-space as the signal is isotropic. However, it is well known that RSDs add anistropies in the 21-cm signal as RSDs introduce distortions along the LoS, and not in directions perpendicular to the LoS. It is, therefore, convenient to bin $\Delta^2_\mathrm{21}$ not only in $k$ but also in $\mu$ where $\mu=\mathrm{cos}(\theta)$ where $\theta$ is the angle between $\mathbf{k}$ and the LoS. 

In Figure~\ref{fig:mu_k_ps} we show how the power spectrum depends on $\mu$ for a selection of wave-numbers (0.1~Mpc$^{-1}$, 0.5~Mpc$^{-1}$, and 1.0~Mpc$^{-1}$ are the dashed green, dotted navy, and solid magenta lines, respectively) and redshifts (20.134, 15.596 and 13.557 are the top, middle, and bottom panels, respectively). We calculated the power spectra for spherical shells with constant k and binned the data into bins of $\mu$. Note that if these curves were calculated from a simulation volume without RSDs, they would be flat. RSD introduces a fourth order polynomial dependence of the power spectrum on $\mu$ \citep[e.g.][]{Barkana2005AFluctuations,Mao2012Redshift-spaceRe-examined}.
At $z=20.134 \Delta^2_\mathrm{21}$ shows a clear dependence on $\mu$, with there being more power for low $\mu$ (i.e. for $\Delta^2_\mathrm{21}$ perpedicular to the LoS). At $z=15.596$, the dependence on $\mu$ has decreased significantly, meaning the magnitude of the anisotropy introduced decreases as heating progresses. This isotropy is due to X-ray heating removing deep absorption from the high density peaks. Finally, the anisotropy has returned by $z=12.603$ when temperature saturation is approached.

It is useful to look in more detail at how the anisotropy evolves with time. We now discuss this using the anisotropy ratio (as defined in \citet{Fialkov2015}), which is given by:
\begin{eqnarray}
r_\mu(k)= \frac{\big< \Delta^2_\mathrm{21}(\mathbf{k})_{|\mu|>0.5} \big>}{\big< \Delta^2_\mathrm{21}(\mathbf{k})_{|\mu|<0.5} \big>} - 1 \,
\end{eqnarray}
where the averages are over angles. 

$r_\mu$ is shown in Figure~\ref{fig:anistropy_ratio} for various wave-numbers (0.1~Mpc$^{-1}$, 0.5~Mpc$^{-1}$ and 1.0 Mpc$^{-1}$ in the top, middle, and bottom panels, respectively). The solid teal lines represent cubes without RSDs (as expected, they are zero), and the dashed maroon lines represent the cubes with RSDs. For the smaller scales shown ($k\sim$ 0.1~Mpc$^{-1}$ and 0.5~Mpc$^{-1}$), the anisotropy decreases as heating commences and increase as temperature saturation approaches. However, for the largest scales considered here ($k\sim$ 0.1~Mpc$^{-1}$), there is a brief increase in the value of $r_\mu$ as heating begins. The f $r_\mu$ increases because the Kaiser effect makes voids seem less dense, so emit a lower value of $\delta T_{\mathrm{b}}$, and heating also causes the value of $\delta T_{\mathrm{b}}$ to decrease while the signal is in absorption. The signal coming from the voids is weak and does not contribute to the small scales. However, the signal from the voids adds anisotropy to large scales as it increases the contrast between the high and low density regions.

\begin{figure*}
  \centering
  \begin{tabular}{c}
    \includegraphics[width=0.67\columnwidth]{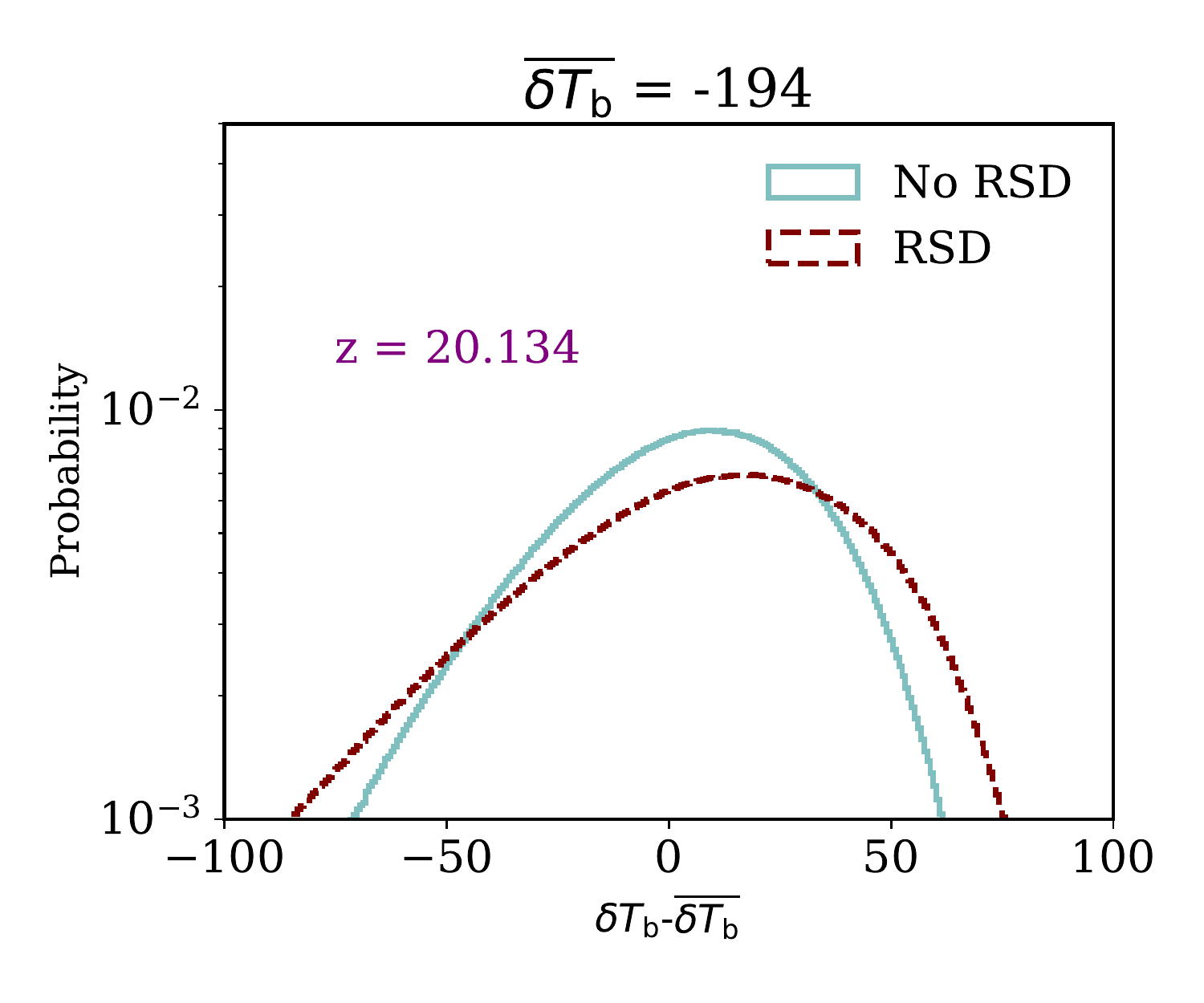} 

    \includegraphics[width=0.67\columnwidth]{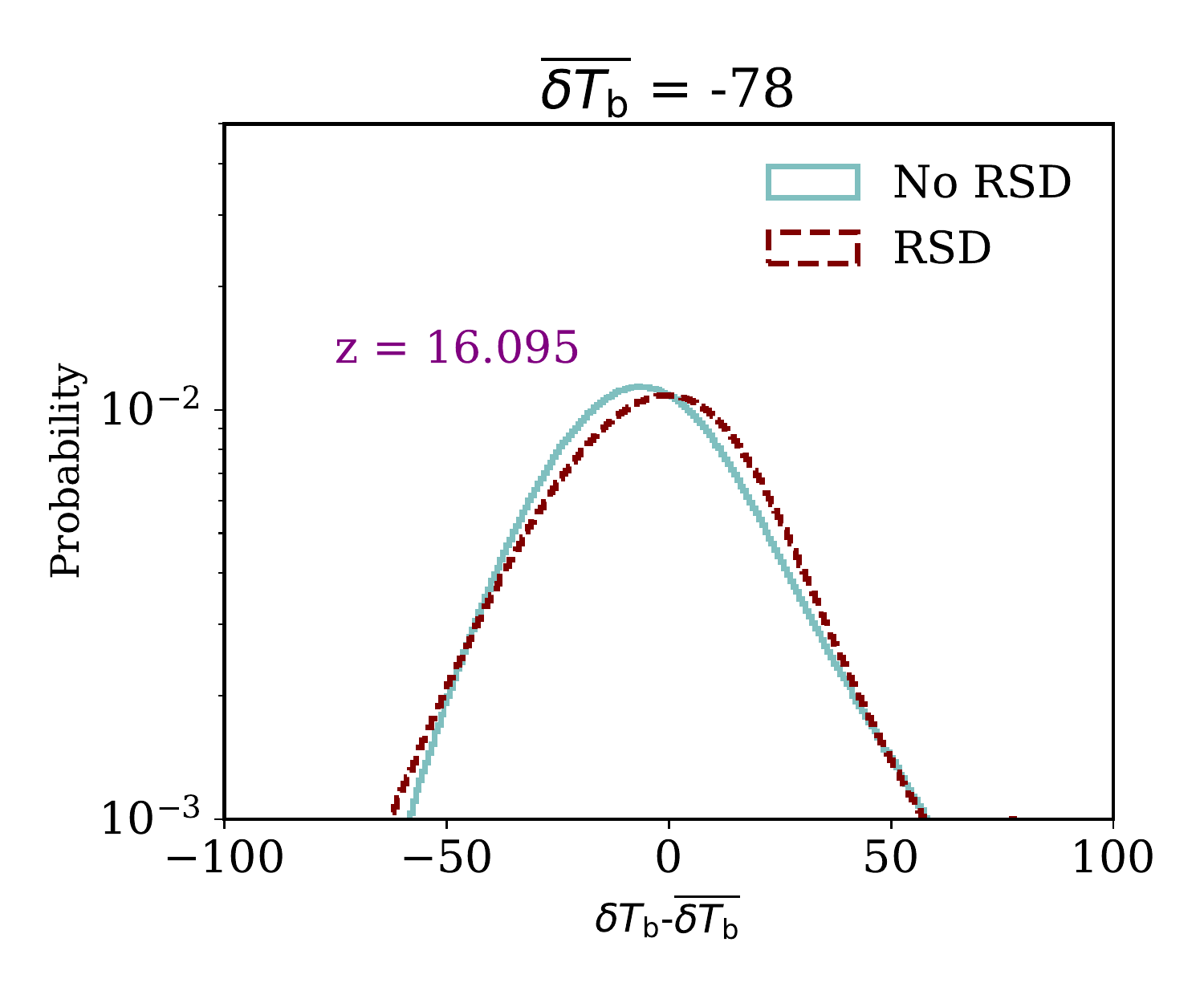} 

    \includegraphics[width=0.67\columnwidth]{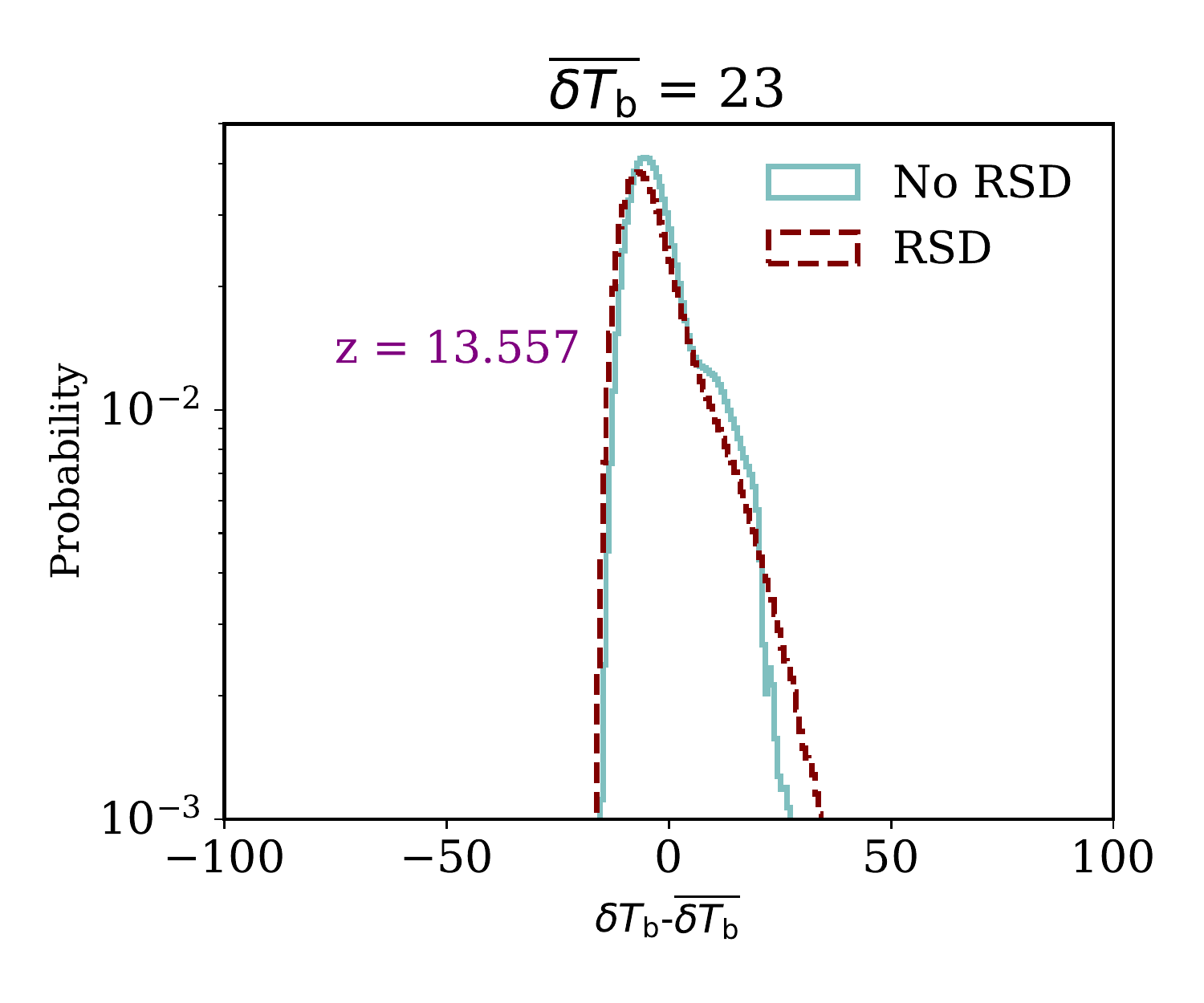} \\
  \end{tabular} \\
  \caption{PDF of the mean subtracted differential brightness temperature over variance at the full simulation resolution with (maroon, dashed) and without (teal, solid) RSDs at a few illustrative redshifts. The kaiser effect is most visible at $z=20.134$ where we can see the PDF has a longer tail on either side, as over-dense regions appear denser and the under-dense regions appear more sparse. While X-ray heating is underway (at $z=16.095$) the PDFs are harder to distinguish thanks to the X-Ray heating having the opposite effect to the kaiser effect while the signal is in absorption. Finally, as temperature saturation is approached at $z=13.557$ the kaiser effect is only visible for the denser regions as the lower end of the PDF is now dominated by partially ionized regions approaching $\delta T_{\mathrm b}=0$.}
  \label{fig:pdfs_rsd}
\end{figure*}

\subsubsection{One-point statistics}

We now move on to discuss the one-point statistics of our simulations. Figure~\ref{fig:pdfs_rsd} shows PDFs of the mean-subtracted signal at simulation resolution. Initially, at $z=20.134$ the inclusion of RSDs cause there to be more points in both deep and shallow absorption as the Kaiser effect causes the dense points to appear denser in redshift-space and the voids more sparse. While the signal is in absorption this corresponds to the lowest (dense areas) and highest (void-like) values of $\delta T_{\mathrm b}$. At $z=16.095$ the transition from absorption to emission has begun and the PDF with RSDs now has fewer points in very deep absorption as X-rays heat first heat the dense regions bringing them into emission. Voids now produce the lowest values of $\delta T_{\mathrm b}$ in the simulation volume as they are furthest from the haloes and are heated last. At $z=13.557$ the signal has transitioned into emission and the kaiser effect is again clearly visible at the high end of the PDF. However, for the lower values of $\delta T_{\mathrm b}$ there is no difference between the two curves due to the onset of reionization. Sources have begun to ionize the high density regions surrounding them, which drives $\delta T_{\mathrm b}$ towards zero.

\begin{figure}
  \centering
  \begin{tabular}{c} 
    \includegraphics[width=1.\columnwidth]{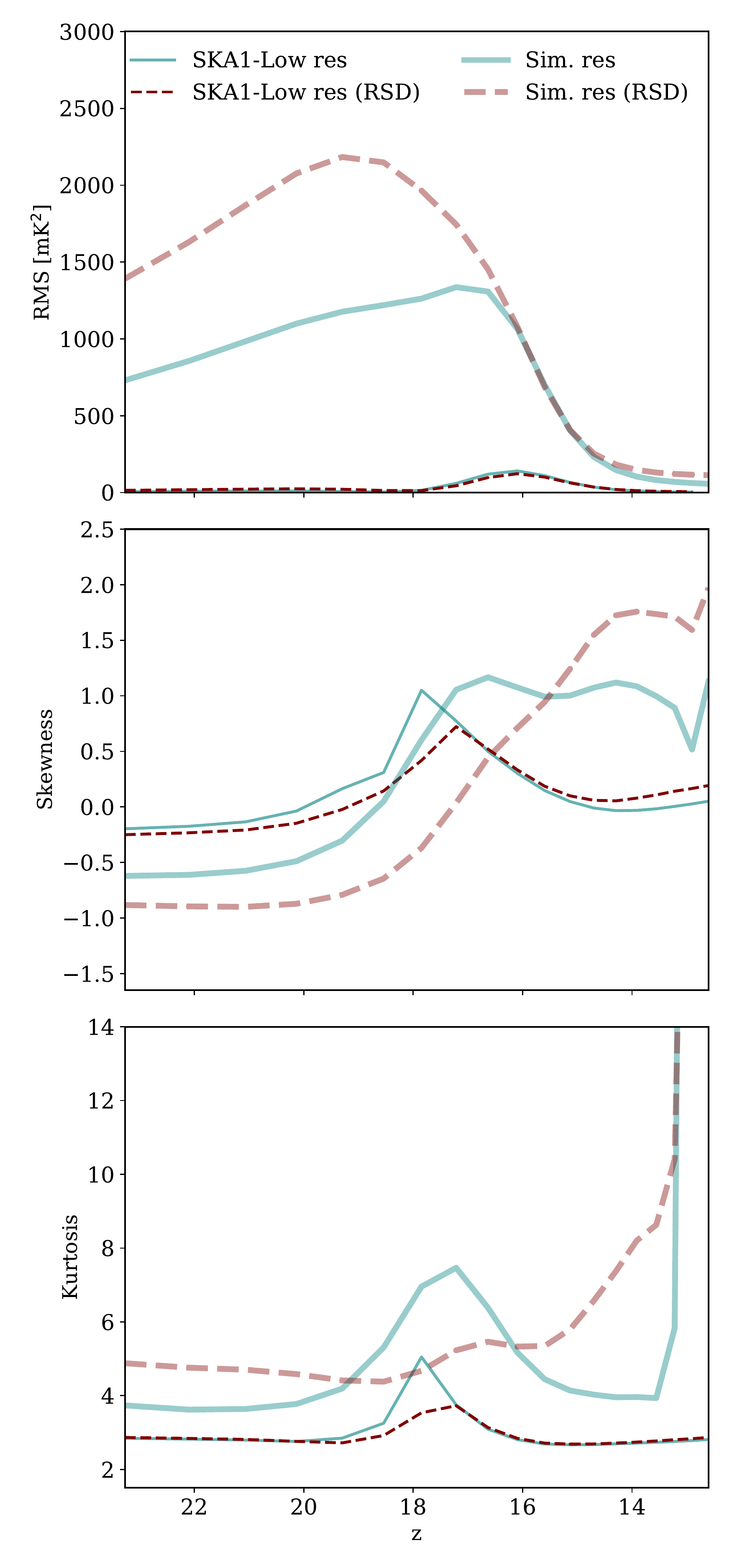} 
  \end{tabular} \\
  \caption{The RMS, skewness and kurtosis for simulations with (dashed red line) and without (solid blue line) RSDs. Thick lines represent the statistics from simulation resolution and thin lines represent the statistics from images smoothed to the expected SKA1-Low resolution.}
  \label{fig:onepoint}
\end{figure}

$\Delta^2_\mathrm{21}$ does not contain any non-Gaussian information about the signal. As the signal from the CD is expected to be highly non-Gaussian \citep[e.g.][]{Watkinson2015TheMoments,Ross2017}, we next consider the skewness and kurtosis.  We use the dimensionless skewness and kurtosis given by:
\begin{equation}
\mathrm{Skewness}(y) = \frac{1}{N} \frac{\sum_{i=0}^{N}(y_i - \overline{y})^3}{\sigma^{3}},
\label{eq:skew}
\end{equation}
and
\begin{equation}
\mathrm{Kurtosis}(y) = \frac{1}{N} \frac{\sum_{i=0}^{N}(y_i - \overline{y})^4}{\sigma^4}.
\label{eq:kur}
\end{equation}
Here, $y$ is the quantity of interest (i.e., $\delta T_{\rm b}$), $N$ is the total number of data points, and $\overline{y}$ and $\sigma^2$ are the mean and variance of $y$, respectively. 

Figure~\ref{fig:onepoint} shows the RMS, skewness, and kurtosis in the top, middle, and lower panels, respectively. The thick lines represent statistics calculated at the simulation resolution, and the thin lines show the same statistics at the expected resolution of SKA1-Low. Again, the red dashed line represents the simulation with RSDs and the solid blue line those without. At the resolution of the simulation, the RMS is boosted significantly by the inclusion of RSDs before and after the time inhomogeneous heating drives the fluctuations (as was seen in the previous plots of $\Delta^2_\mathrm{21}$), and peaks at a value of around 2200~mK$^2$ as opposed to about 1400~mK$^2$ when RSDs are not included. However, at the expected resolution of SKA1-Low the effect is almost unnoticeable. 

The skewness in the middle panel of Figure~\ref{fig:onepoint} is initially about two thirds lower (i.e. a magnitude two thirds higher) in redshift-space than real-space, as initially RSDs skew the PDF to the more negative values of $\delta T_{\mathrm{b}}$ as high density points emit $\delta T_{\mathrm{b}}$ in deep absorption. As the signal transitions to emission the PDF for the simulation with RSDs starts to move toward more positive values (as high density points now emit a $\delta T_{\mathrm{b}}$ signal in strong emission), resulting in a value approximately 2 times higher for the skewness in redshift-space. However, this effect appears to be less obvious at SKA1-Low resolution.

The kurtosis (shown in the lower panel of Figure~\ref{fig:onepoint}) is a measure of the `tailedness' of the PDF. We first consider the kurtosis from the simulation resolution, represented by the thicker lines.  Initially, the simulation with RSDs has a greater kurtosis (of approximately 5 compared to around 3.7 at simulation volume) as the high density regions give more negative values.  However, the most high density regions are quickly heated as they are closest to the sources, which removes this initial negative $\delta T_{\mathrm{b}}$ tail. While X-ray heating is underway (and before temperature saturation is reached) the  simulation without RSDs shows a large increase in non-Gaussianity peaking at around 7.5. This peak is far smaller in the kurtosis from the simulation when RSDs are included, only reaching approximately 5.5. The fluctuations responsible for this increased non-Gaussianity are large-scale heating fluctuations, and they are also the fluctuations seen to be slightly suppressed in $\Delta^2_\mathrm{21}$. As the highest density regions reach emission at which point they form another tail at the positive $\delta T_{\mathrm{b}}$ region, an effect that is enhanced by the kaiser effect, resulting in an increase in the kurtosis in the model with RSDs. A dramatic increase in kurtosis occurs at the very end of the simulation for both models, due to high density regions becoming ionized as reionization begins, resulting in the decrease of $\delta T_{\mathrm{b}}$ adding another tail. The increase in kurtosis is almost entirely lost when the simulation is smoothed to SKA1-Low resolution, presumably because the high density points are not resolved.

At SKA1-Low resolution (represented in the lower panel of Figure~\ref{fig:onepoint} by the thin lines) the peak in kurtosis from X-ray heating is still present at SKA1-Low resolution, with a far smaller magnitude. Again, the peak value is higher when RSDs are not included (approximately 5.1) than when they are (approximately 3.6). Unfortunately this suggests that the kurtosis will not be as effective at probing the timing of X-ray heating as previously thought. Other than the peak driven by X-ray heating the two models produce the same value of kurtosis, which remains approximately constant at around 3.0 for the rest of the simulation.

\subsection{Differences between the different astrophysics models}
\label{sec:allmodels}

In this section, we will consider the differences between our astrophysical models and their detectability. The parameters of the different sources in each model are summarized in Table~\ref{tab:models}. Firstly, we will summarize the previous findings for these models. Models with early Ly-$\alpha$ saturation (ending in `-EL') assume, as with the fiducial model, that Ly-$\alpha$ saturation has been reached before the simulation begins. Models with late Ly-$\alpha$ saturation (ending with `-LL'), however, assume that the sources in our simulations are the first, ensuring the signal is completely coupled to the CMB initially. In these late Ly-$\alpha$ coupling scenarios, the Ly-$\alpha$ background is an additional source of fluctuations until Ly-$\alpha$ saturation is reached at approximately $z\sim16$. Models beginning with `S1-' and `S2-' contain HMXBs result in relatively early heating when compared to the other models (beginning with `S3-' and `S4-'), which only contain QSOs that yield a much later heating scenario. The QSOs with a softer spectra (in the `S4-' models) heat the IGM slightly less rapidly as they produce lower energy photons. As QSOs are rarer sources, they increase the non-Gaussianity of the signal produced, so all models that contain them have higher values of skewness and kurtosis (`S2-',`S3-' and `S4-').

\begin{figure*}
  \centering
  \begin{tabular}{c}
    \includegraphics[width=1.\columnwidth]{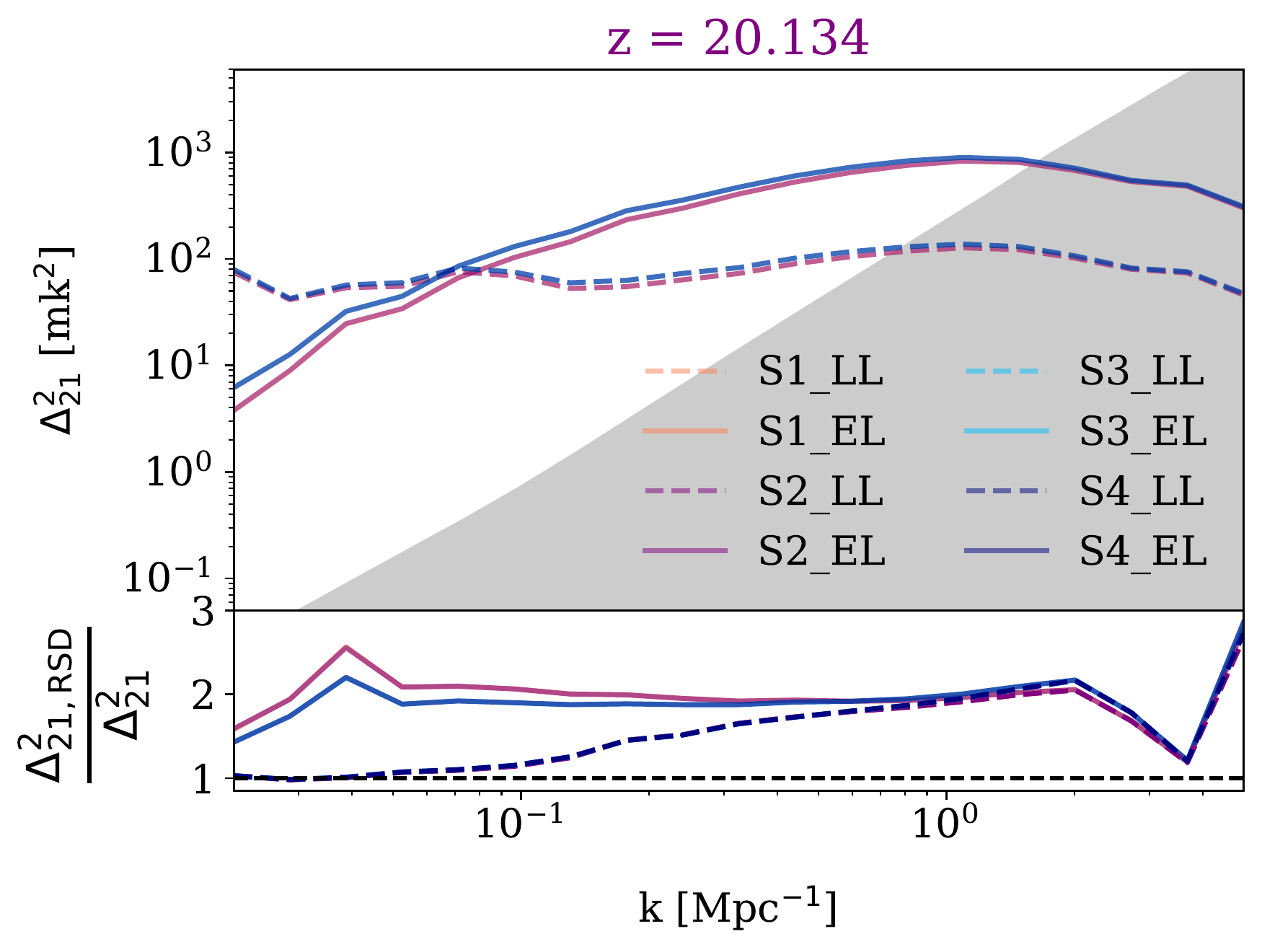}
    \includegraphics[width=1.\columnwidth]{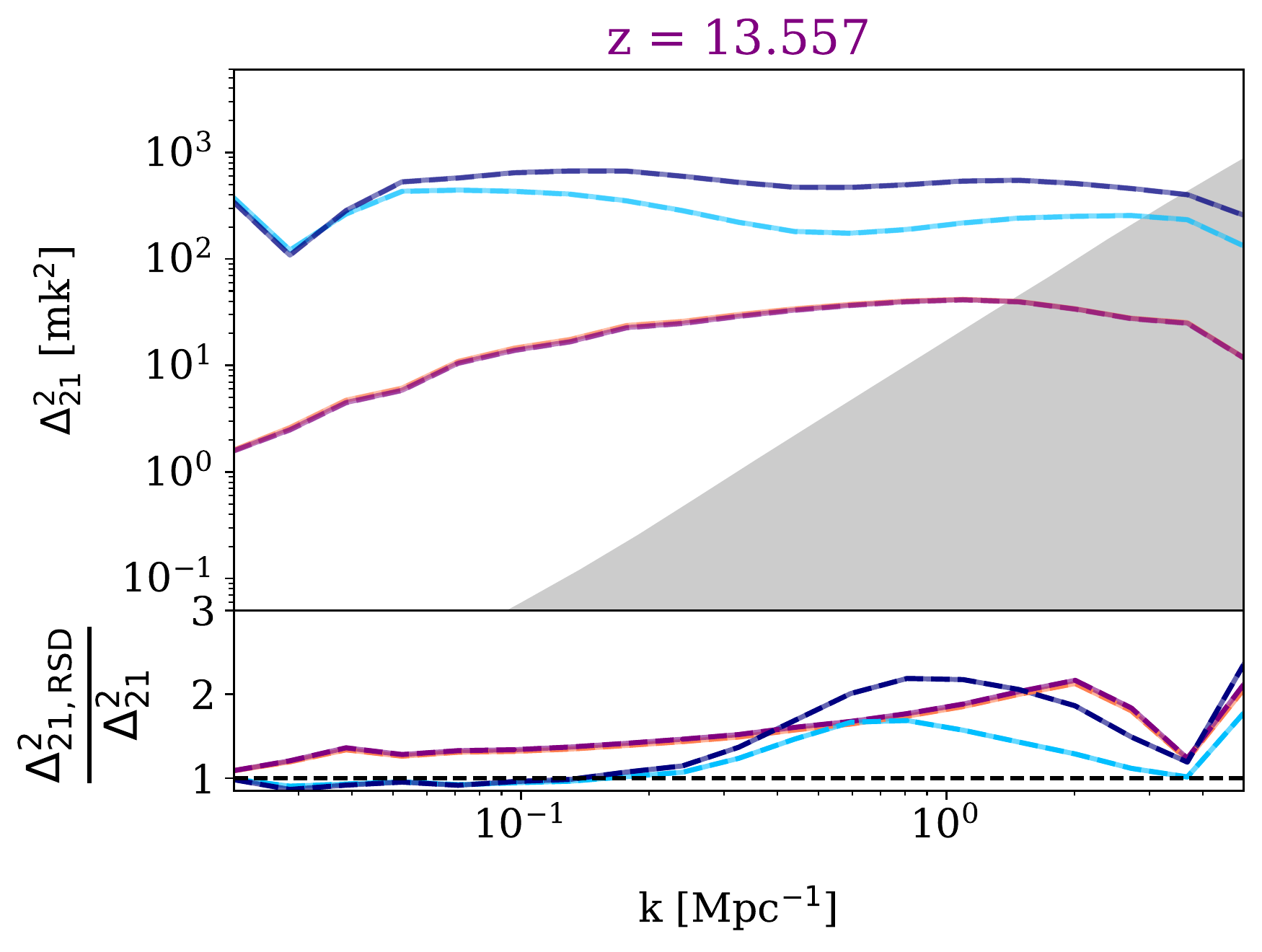} \\   
    
    \includegraphics[width=1.\columnwidth]{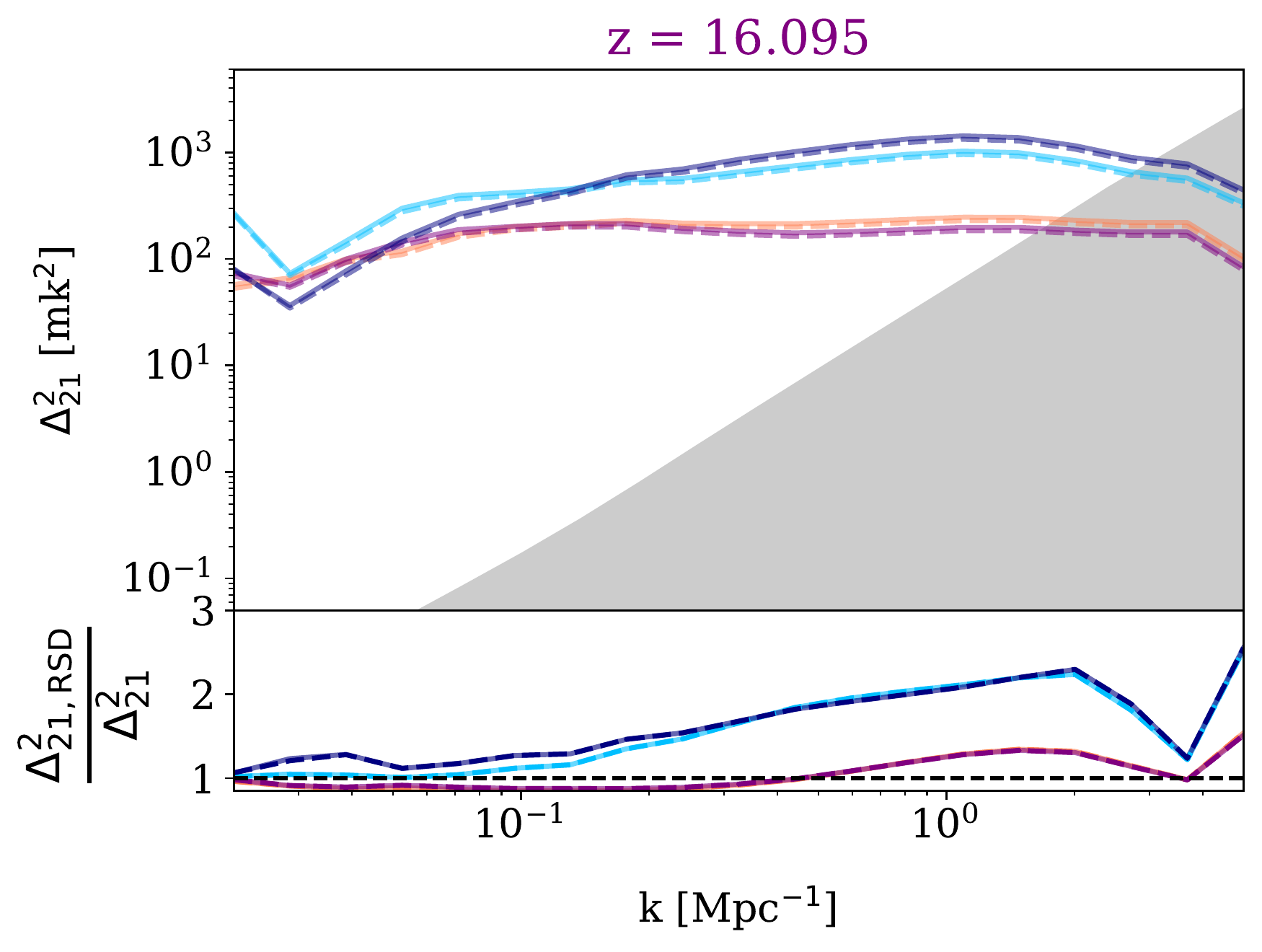} 
    \includegraphics[width=1.\columnwidth]{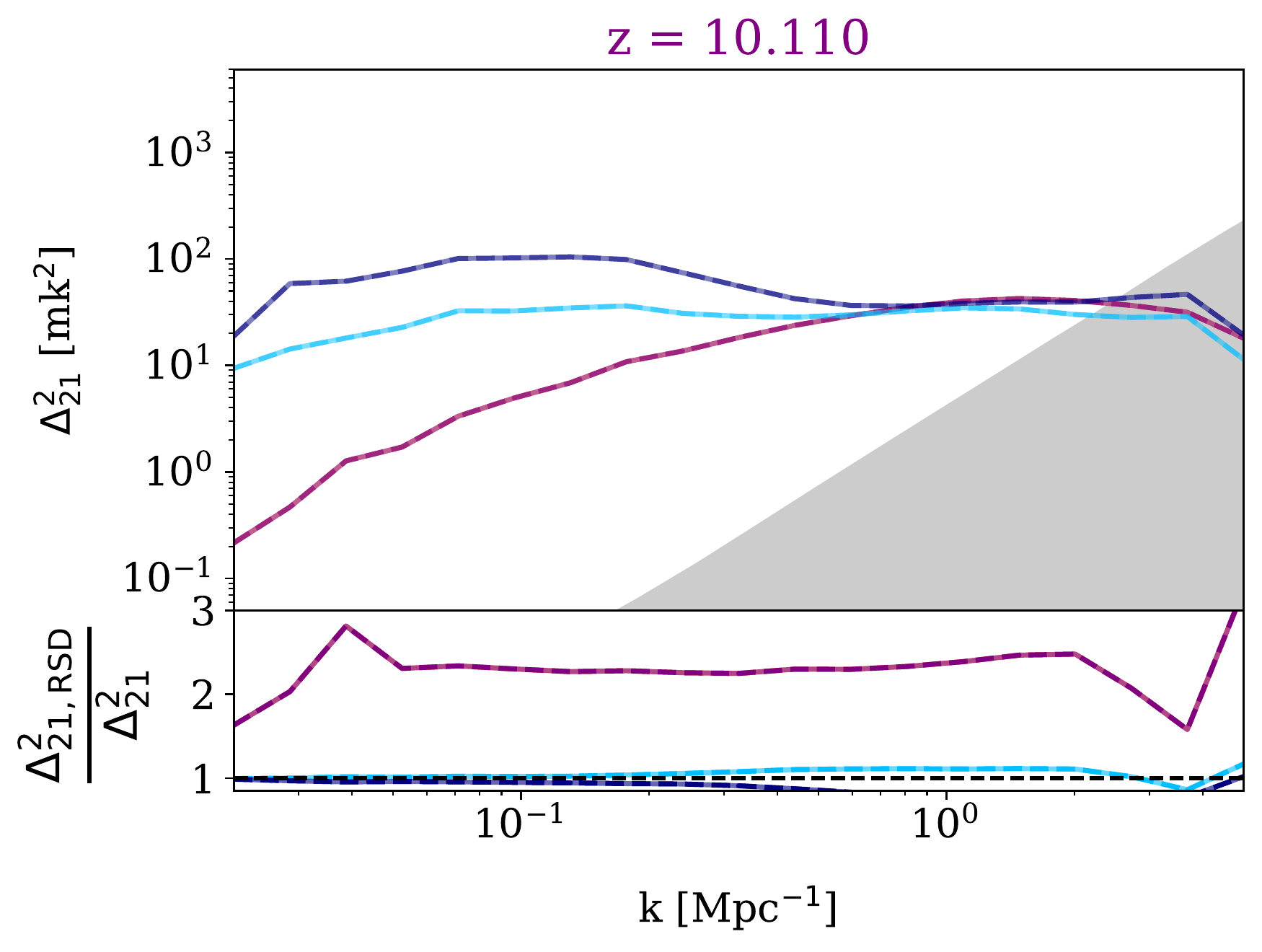} \\
    
    \includegraphics[width=1.\columnwidth]{plots_allS/ps_all_models_z13.557.pdf} 
    \includegraphics[width=1.\columnwidth]{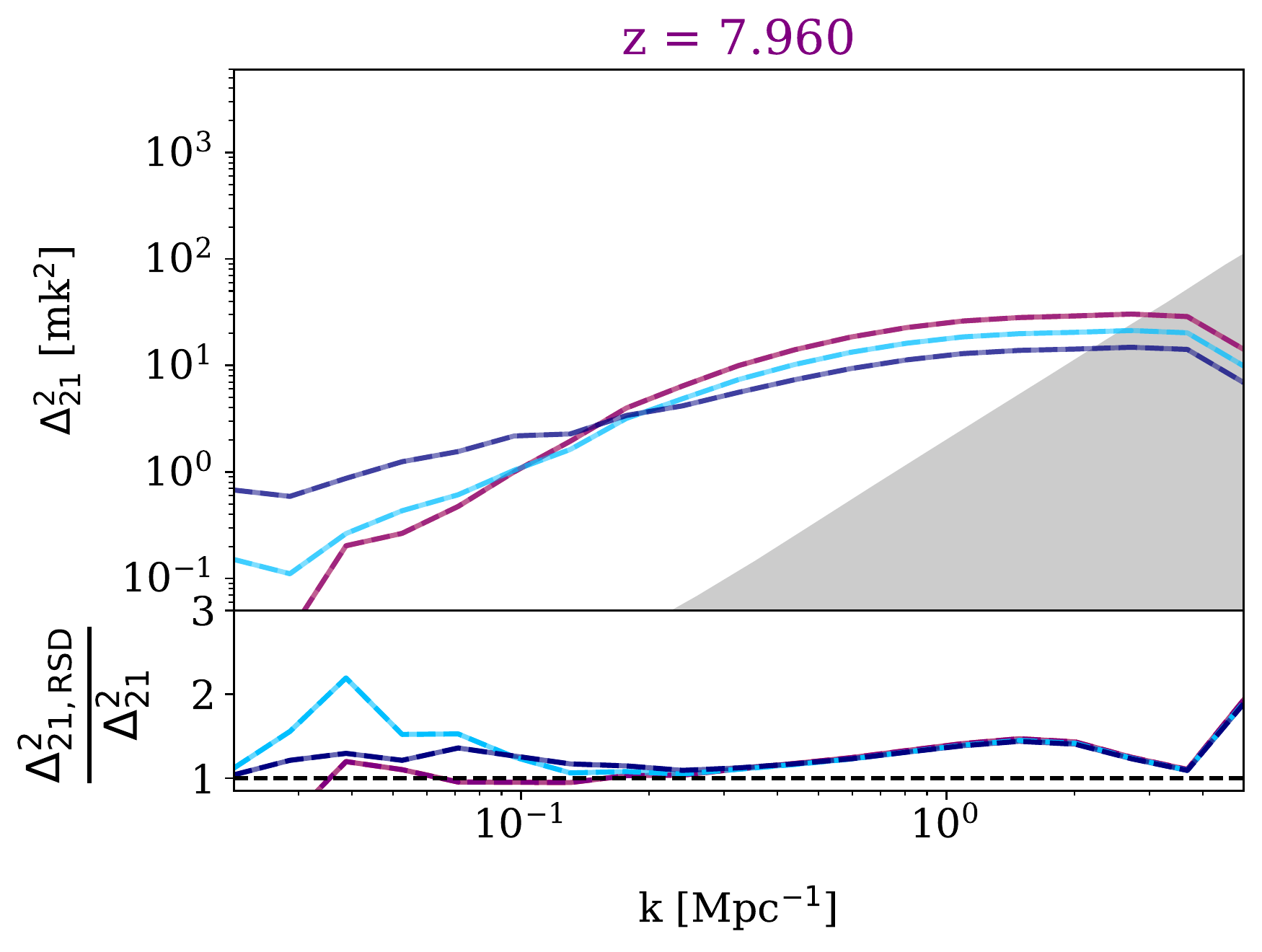} \\
  \end{tabular} \\
  \caption{$\Delta^2_\mathrm{21}$ for all astrophysical models with RSDs included is shown in the upper panel of each plot. The shaded grey area shows the expected telescope noise for SKA1-Low. In the lower panels we show the ratio of $\Delta^2_\mathrm{21}$ with to $\Delta^2_\mathrm{21}$ without RSDs. For late Ly-$\alpha$ saturation models $\Delta^2_\mathrm{21}$ is greater in redshift-space than in real-space thanks to the in-homogeneous Ly-$\alpha$ background. As was found in the previous section $\Delta^2_\mathrm{21}$ is greater for all models in redshift-space than real-space before significant heating occurs and after temperature saturation has been reached, although the timing of this varies between different models.}
  \label{fig:all_ps}
\end{figure*}
Figure~\ref{fig:all_ps} displays $\Delta^2_\mathrm{21}$ for all astrophysical models in the upper half of each panel. While the late Ly-$\alpha$ models are more likely to produce a $\Delta^2_\mathrm{21}$ measurement for very high wave-numbers  ($k<0.06$~Mpc$^{-1}$) at the beginning of the CD ($z\sim20$), they are less detectable for all wave-numbers greater than this. The early models are above the expected noise from our model for $k<0.2$ whereas the late models are above it for a slightly larger range, $k<0.8$. From the ratio plots, it can be seen that the inclusion of RSDs boosts the power on scales where the power is close to the noise level expected from our SKA1-Low noise model. Thus, the likelihood of making a $\Delta^2_\mathrm{21}$ measurement with this SKA1-Low noise scenario is greater than previously anticipated (i.e.\ before RSDs were added to the simulations) for all models. However, the detectability of the models with later heating is not as significantly increased by the inclusion of RSDs, as the signal from the high density peaks is removed by ionization before temperature saturation is reached. However, these later heating models still produce $\Delta^2_\mathrm{21}$ with the greatest magnitude, as was previously found in \citet{Ross2019EvaluatingDawn}.

From the ratio plots on the lower half of the panels in Figure~\ref{fig:all_ps}, we can see that the late (inhomogeneous) Ly-$\alpha$ background (dashed lines) removes the boost in power from the RSDs on large scales. All models with early Ly-$\alpha$ saturation (solid lines) start with a ratio of about $1.87$, which decreases as heating begins. As was found previously \citep{Ross2019EvaluatingDawn}, models with an inhomogeneous Ly-$\alpha$ background exhibit an initial suppression in power on all scales. Heating progresses differently for varying models, causing the ratio to decrease more slowly in the models with later heating (S3~LL, S3~EL, S4~LL, S4~EL). While the earlier heating models see the ratio increase to around $1.87$ once more before reionization begins, the later heating models never approach this value again as reionization has started before temperature saturation is completed. 

\begin{figure*}
\centering
  \includegraphics[width=2\columnwidth]{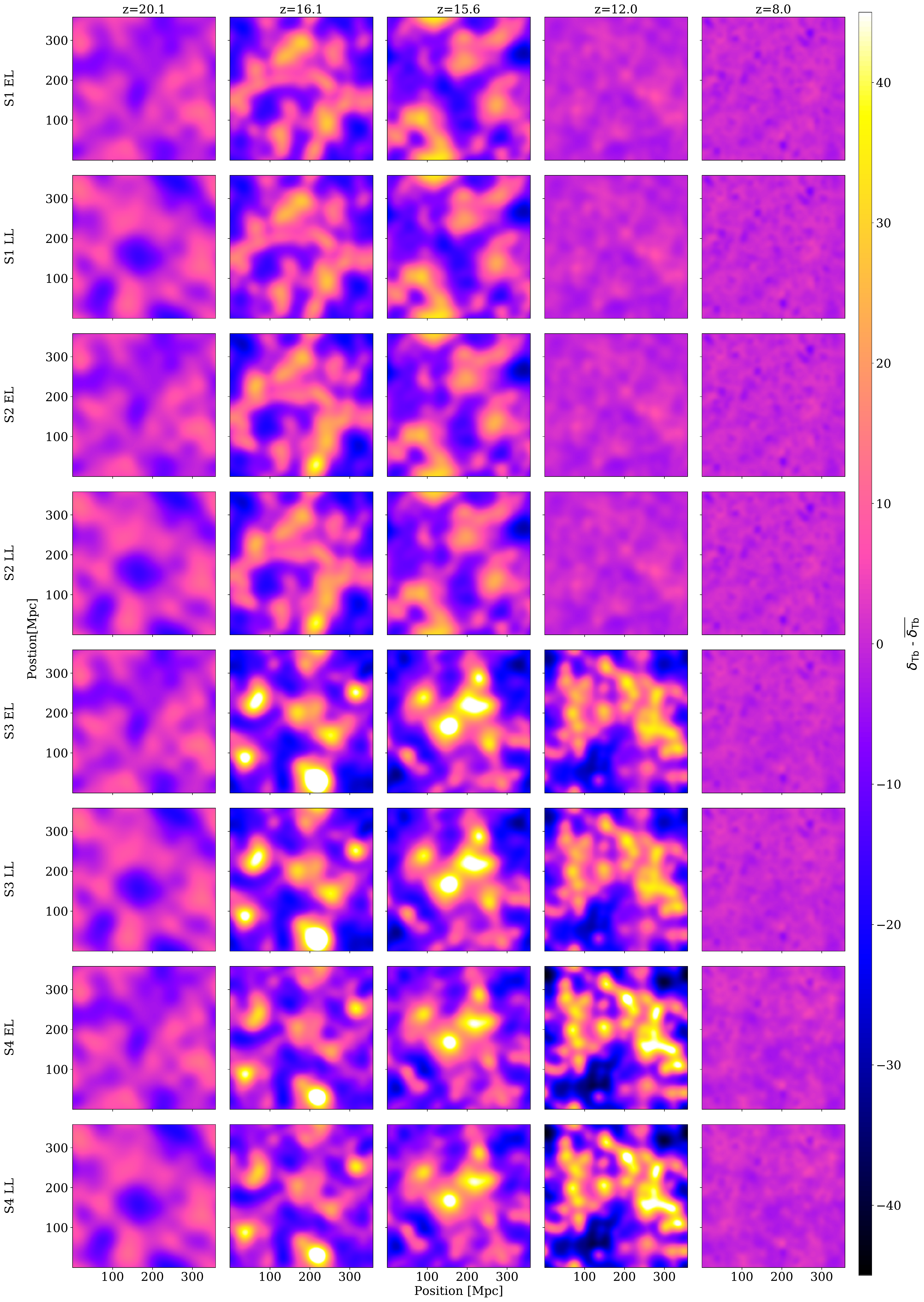} 
  \caption{Images of the mean subtracted 21-cm differential brightness temperature for all astrophysical models at the expected SKA1-Low resolution.}
  \label{fig:SKAres_images}
\end{figure*}

\begin{figure*}
\centering
  \includegraphics[width=2\columnwidth]{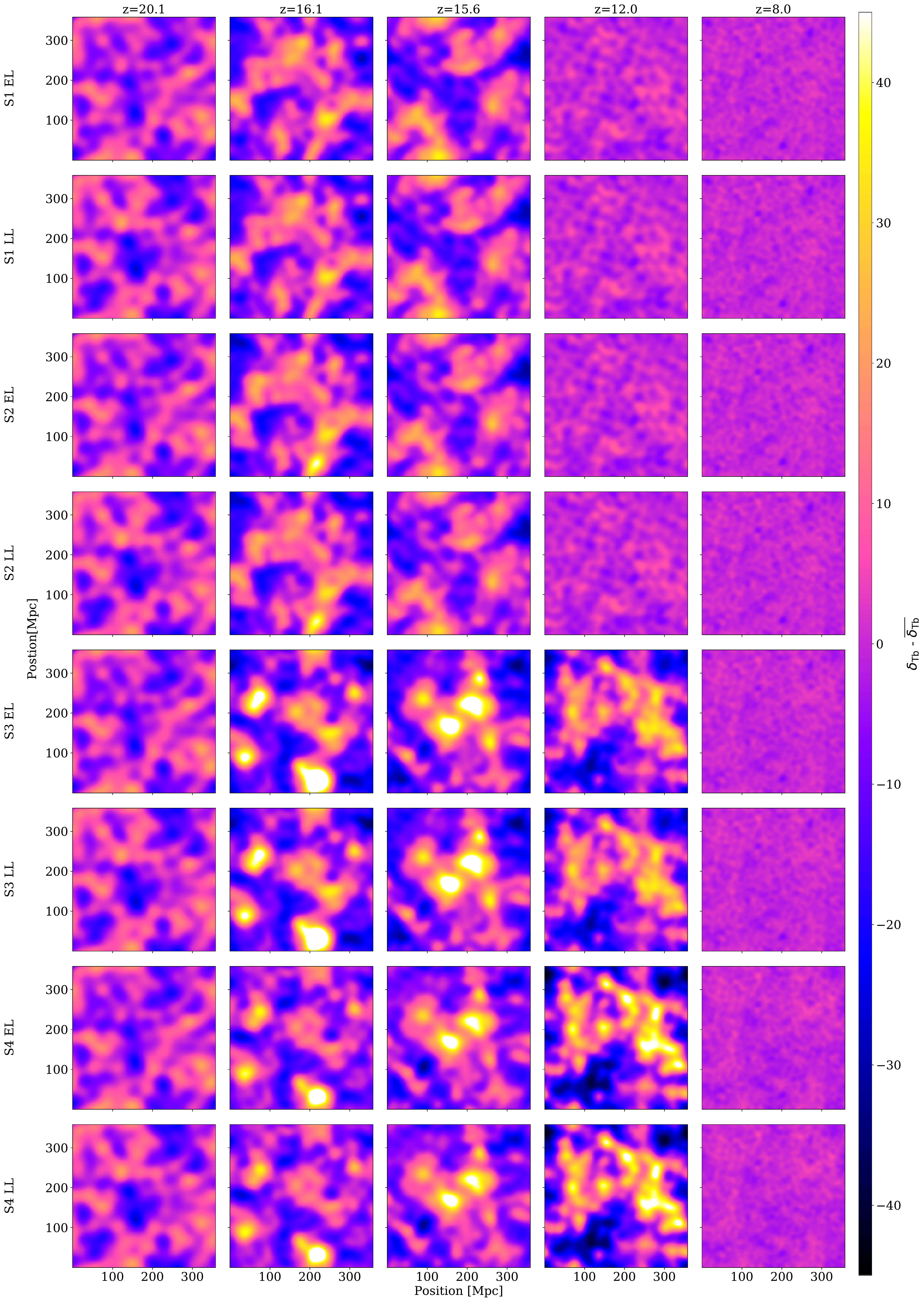}
  \caption{The same as Figure~\ref{fig:SKAres_images} but this time with the expected SKA1-Low resolution, this time with telescope noise from Section~\ref{sec:noise} added.
  }
  \label{fig:SKAres_images_noise}
\end{figure*}

Figure~\ref{fig:SKAres_images} shows mock images from our suite of simulations at the expected resolution of SKA1-Low (1.2 km maximum baseline). We degrade our images to the resolution expected by SKA1-Low. Images have been smoothed with a Gaussian beam with a FWHM corresponding to a 1.2 km maximum baseline at the relevant frequency and bandwidth-smoothed with a top-hat function (width equal to the distance corresponding to the beam width). As was found previously in \citet{Ross2019EvaluatingDawn}, all images are distinguishable by eye at SKA1-Low resolution. The difference between late and early Ly-$\alpha$ saturation is also visible at this resolution, which was not shown previously.

In Figure~\ref{fig:SKAres_images_noise}, we add simulated telescope noise (as described in Section~\ref{sec:noise}) to our mock images. The features introduced to the signal from the long X-ray heating are clearly visible even once this noise has been added. In addition, a difference between early and late Lyman-alpha saturation is also visible with telescope noise. This difference further strengthens our previous claim that the imaging in the CD will be possible with SKA. The reader should note that the telescope noise considered in this work is quite optimistic. The slices in Figure~\ref{fig:SKAres_images_noise} tells us that the signal high enough for the the features to be visible in the images even when the noise increases by a factor of $\sim 5$. However, we have not considered the foregrounds from 21-cm observations. The residual foregrounds can make imaging of the CD difficult. Therefore, one should consider these mock 21-cm images as the most optimistic case.

\begin{figure*}
  \centering
  \begin{tabular}{c}
          \includegraphics[height=1.5\columnwidth]{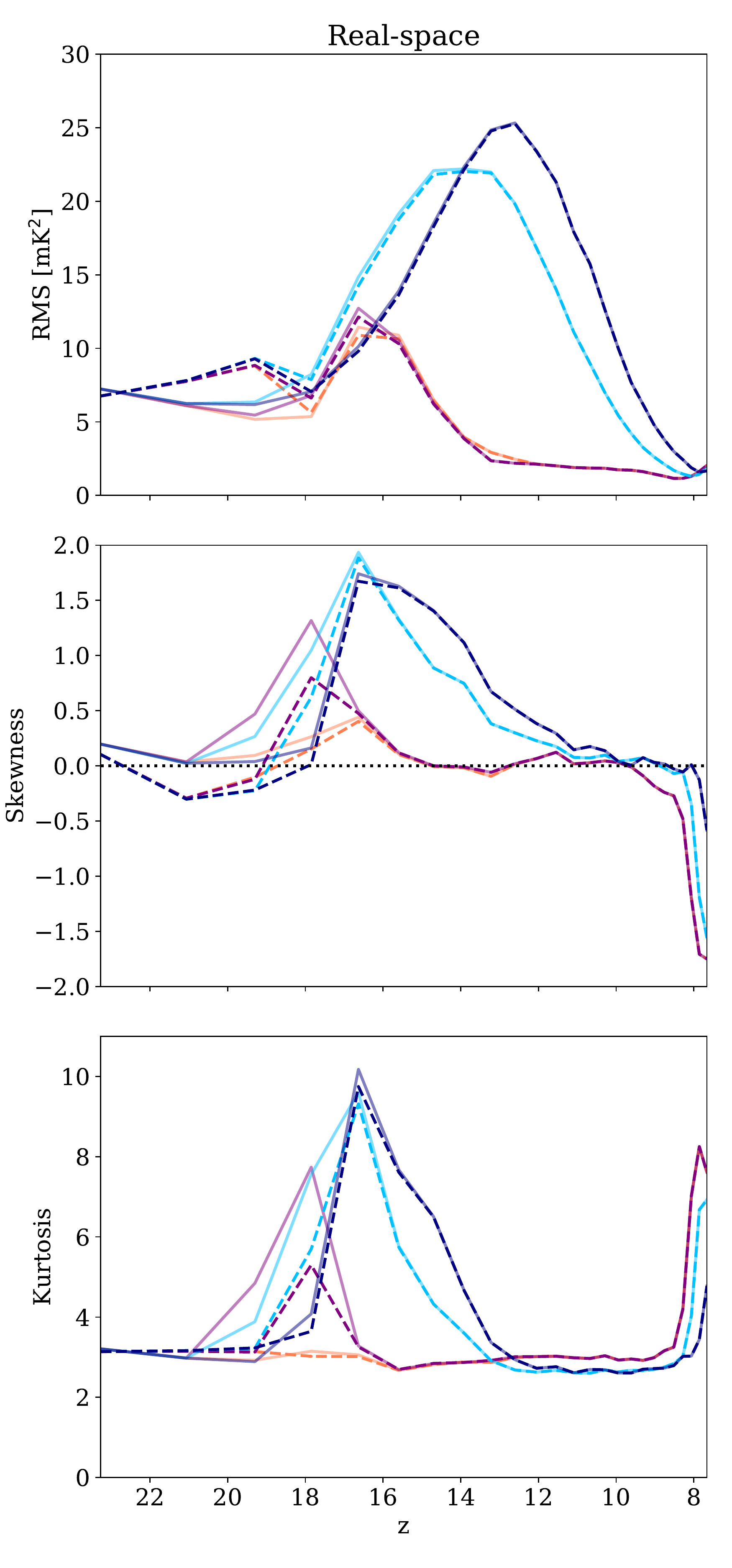} 
          \includegraphics[height=1.5\columnwidth]{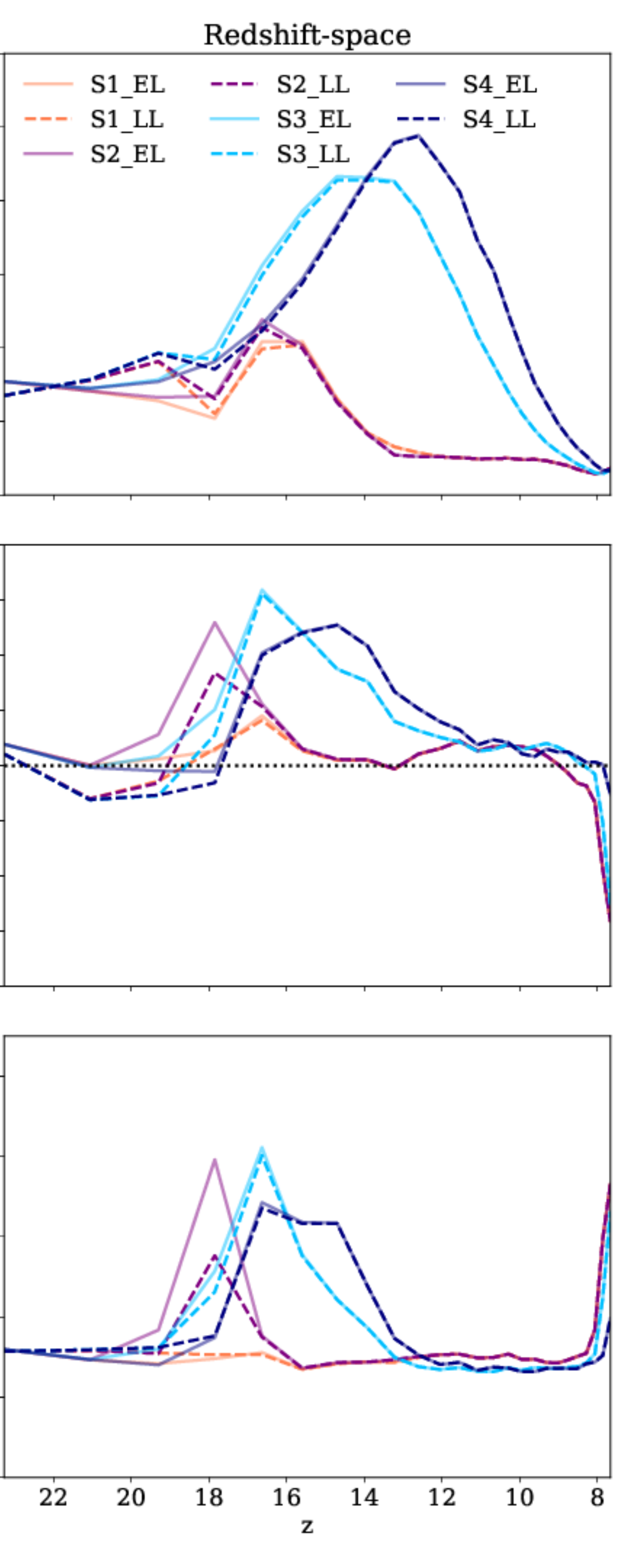} 
          \includegraphics[height=1.5\columnwidth]{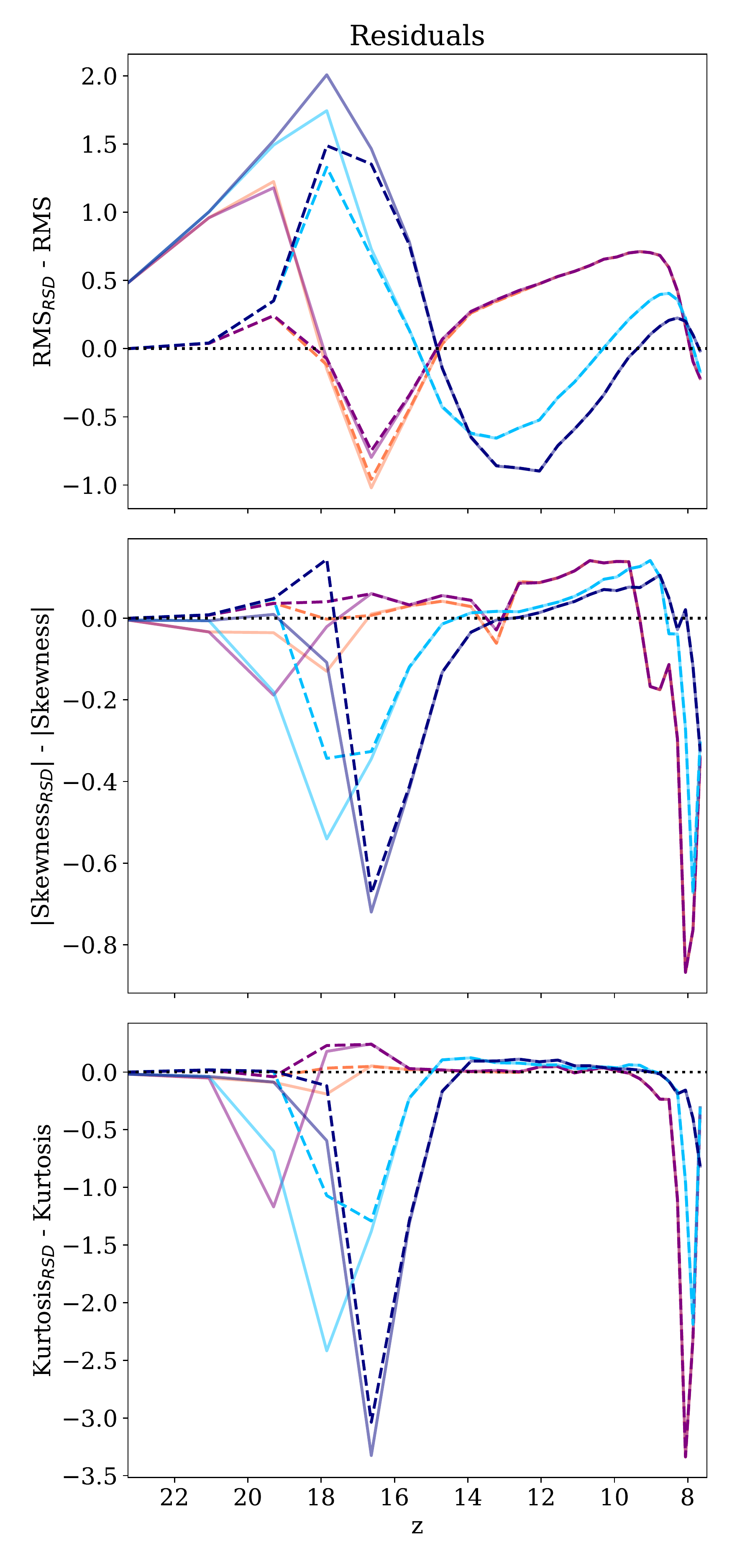} 
  \end{tabular} 
  \caption{RMS (top panels), skewness (central panels), and kurtosis (lowest panels) at the expected resolution of SKA1-Low with noise from our model added for all models both without (left-hand panels) and with (centre panels) RSDs. The residuals between the statistics in redshift-space and real-space are shown in the right-hand panels. The reader should note that we have shown the residuals of the magnitude of the skewness for simplicity. The RMS from models with a homogeneous Ly-$\alpha$ background (early models, solid lines) are unchanged at early times. However, the in-homogeneous Ly-$\alpha$ background in late Ly-$\alpha$ models (dashed lines) gives a boost to the RMS at high redshift. The magnitude of the skewness and kurtosis are almost always decreased by the inclusion of RSDs.}
  \label{fig:onepoinst_allmodels}
\end{figure*}

Finally, Figure~\ref{fig:onepoinst_allmodels} shows the RMS, skewness, and kurtosis in the upper, middle, and lower rows, respectively, for all astrophysical models at SKA1-Low resolution with telescope noise included. The left-hand column shows these statistics in real-space, the central column in redshift-space, and the residuals are plotted in the right-hand-side column. We have chosen to show residuals rather than ratios as the ratios are rather noisy due to some of the values, particularly the skewness, being very low. Again, it should be noted that foregrounds have not been taken into account, making this a best-case scenario. 

As found previously, all the models produce distinguishable RMS histories, although the early heating models (S1-El, S1-LL, S2-EL and S2-LL) follow a somewhat similar evolution as the HMXBs dominate X-ray heating. At early times the RMS shows a clearly different evolution for the different Ly-$\alpha$, early Ly-$\alpha$ saturation (solid lines) and late Ly-$\alpha$ saturation (dashed lines), both with and without RSDs. The residuals show a boosted RMS (up to 2~mK$^2$ higher for the late X-ray heating models and 1~mK$^2$ for early models) from the late Ly-$\alpha$ saturation models in redshift-space at early times. Conversely a decrease in RMS when X-ray heating dominates (up to approximately 1~mK$^2$ lower for all models). This behaviour is due to the fact that an inhomogeneous Ly-$\alpha$ background tends to cause the low-density regions to be closer to zero (as they are not yet coupled, being far away from sources), and X-ray heating tends to do the opposite (as regions closer to the sources tend to be heated first, which decreases their magnitude). Therefore, as discussed previously, an inhomogeneous Ly-$\alpha$ background builds on the Kaiser effect, rather than cancelling it out as X-ray heating does.

Overall, the skewness (middle panels of Figure~\ref{fig:onepoinst_allmodels}) and kurtosis (lower panels) follow a similar evolution to that found in our previous work \cite{Ross2019EvaluatingDawn} generally with slightly decreased magnitudes. This decrease in non-Gaussianity is due to fluctuations from X-ray heating and the kaiser effect effectively cancelling each other out. The skewness and kurtosis both show that the non-Gaussianity is greater in redshift-space after temperature saturation has been reached, as is to be expected as the signal is now dominated by density fluctuations. Finally, the skewness and kurtosis show that the non-Gaussianity in redshift-space is less than that in real-space at the onset of reionization. However, the skewness and kurtosis remain good metrics to distinguish between the models with HMXBs and both HMXBs and QSOs that produce similar power spectra and RMS measurements. 

Late Ly-$\alpha$ saturation (solid lines) models yield a larger negative skewness in redshift space, whereas early Ly-$\alpha$ saturation (dashed lines) do not show a significant difference. This difference between the two Ly-$\alpha$ scenarios is due to the fact that late Ly-$\alpha$ coupling means there are many points close to zero where the signal has not yet decoupled from the CMB, which skews the signal towards larger values, than in the early Ly-$\alpha$ scenario. In previous work without RSDs, this effect was less pronounced. The Late Ly-$\alpha$ scenario produces a higher kurtosis than the early one. This difference was not previously visible at SKA1-Low resolution with telescope noise and is caused by the inhomogeneous Ly-$\alpha$ background, resulting in much of the signal being close to zero and those points that have been coupled forming a tail. At the end of the simulations, as reionization begins, the inclusion of RSDs appears to decrease the magnitude of both the value of the skewness and kurtosis in redshift-space compared to real-space.  

\section{Summary}
\label{sec:conclusions}

As we will observe the 21-cm signal in redshift-space, not real-space, it is vital that we include RSDs in our predictions. In this paper, we presented our analysis of the 21-cm signal with and without RSDs from our suite of fully numerical RT simulations presented in \citet{Ross2017, Ross2018} and \citet{Ross2019EvaluatingDawn}. We first showed detailed statistics for our fiducial model, emboldened in Table~\ref{tab:models}, including both the isotropic and anisotropic power spectra, evolution of various wave-numbers and the the anisotropy ratio with redshift, and the one-point statistics (PDFs, RMS, skewness, and kurtosis). We commented on the differences in this more realistic 21-cm signal now that we have included RSDs and its detectability. We then showed the differences between our different astrophysical models with the inclusion of RSDs and demonstrated that they remain distinguishable and detectable.

RSDs boosted fluctuations when the fluctuations in the differential brightness temperature are dominated by the density field due to the Kaiser effect. Initially, the ratio of $\Delta^2_\mathrm{21}$ on large scales matched the amplification factor of 1.87 expected for the matter power spectra \citep{Mao2012Redshift-spaceRe-examined}, but decreased as the simulation continued. This behaviour is somewhat similar to that found in \citet{Jensen2013ProbingDistortions}, with the boost from RSDs decreasing as the magnitude of the 21-cm signal from the densest points decreased (in the case of the EoR, this was due to ionized regions decreasing the value of $\delta T_{\mathrm{b}}$, in the CD this was because of X-ray heating increasing the value of $\delta T_{\mathrm{b}}$). However, unlike in the EoR, the boost in power did not continue to decrease, but increased once more as temperature saturation was approached. In the CD the decrease in the ratio of $\Delta^2_\mathrm{21}$ is caused by dense regions to having a high magnitude value while the signal is in absorption, whereas heating preferentially raises the value of $\delta T_{\mathrm b}$ in these areas. The X-ray heating does not significantly impact the voids at early times as much as the dense peaks so it doesn't significantly interfere with the kaiser effect there (where it would behave in the same way, raising $\delta T_{\mathrm b}$). We saw a similar effect when our in-homogenous lyman-alpha background overlapped with the onset of X-ray heating in \citep{Ross2019EvaluatingDawn}: rather than the fluctuations adding to each other they cancelled each other out. Once X-ray heating starts affecting the voids, the signal has transitioned to emission. Now the X-ray heating does the opposite to the Kaiser effect in these areas. The kaiser effect makes the voids more voidy, which now decreases the value of $\delta T_{\mathrm b}$, but X-ray heating increases it. While the boost in power was often lower than 1.87 and never reached the upper limit of $\sim$5 set by \citet{Mao2012Redshift-spaceRe-examined,Ghara201521cmVelocities}, we showed that the inclusion of RSDs slightly increases the detectability of $\Delta^2_\mathrm{21}$ for all redshifts for all our other astrophysical models. While small, this boost further strengthened hopes that a $\Delta^2_\mathrm{21}$ of the 21-cm signal from the CD may be possible with SKA, even after foreground removal. 

The inclusion of RSDs added anisotropy to the signal, as has been found in many previous works \citep[e.g.][]{Bharadwaj2004TheReionization,Barkana2005AFluctuations,McQuinn2006CosmologicalReionization, Mao2012Redshift-spaceRe-examined, Jensen2013ProbingDistortions, Majumdar13, Majumdar2016}. This additional anisotropy appeared as peculiar velocities compressed densities along the LoS in redshift-space. We used the anisotropic power spectrum and anistropy ratio to investigate the impact of RSDs on the signal. Our results showed that while RSDs add anistropies to the signal before X-ray heating has started and once temperature saturation is approaching, these anistropies decreased as X-ray heating progressed. Our results are in agreement with \citet{Jensen2013ProbingDistortions}, as the power spectrum dependence we found at the end of our simulation matched that found in the beginning of their simulations. The behaviour of the anisotropy was found to be very different than what happens to the anisotropy during reionization. The results found in \citet{Jensen2013ProbingDistortions} where the anistropies are not diminished, but the dependence on $\mu$ inverted as reionization progresses. Anisotropy in the EoR was due to the fact that ionization happened locally, so quickly cut out the high-density peaks, whereas in the CD, X-ray heating impacted the whole simulation volume as well as the power spectra on all scales. The signal from the time period that produces the greatest magnitude of fluctuations in the CD appeared to be reasonably isotropic. Therefore, while the anisotropic power spectra was found previously to be useful during the EoR, we found it was unlikely to produce much additional information during the CD. 

The behaviour of the PDFs was somewhat similar to that found in \citet{Jensen2013ProbingDistortions}, with the boost from RSDs decreasing as the magnitude of the 21-cm signal from the densest points decreased (in the case, of the EoR this was due to ionized regions decreasing the value of $\delta T_{\mathrm{b}}$, in the CD this was because of X-ray heating increasing the value of $\delta T_{\mathrm{b}}$). However, the evolution of the PDFs during the CD was complicated by the fact the signal transitioned from absorption to emission. Initially, there were more points in emission, then more points in absorption. This effect also lead to the skewness switching signs, with the skewness from the simulation including RSDs having a greater magnitude at all times compared to the signal without RSDs. We showed mock images from our simulations at SKA1-Low telescope resolution both without and with telescope noise. We found that, even with the addition of telescope noise, features from all models were clearly visible (as was previously predicted in \citet{Ross2019EvaluatingDawn}). However, the RMS of the signal from our fiducial model with and without RSDs included was indistinguishable suggesting imaging the signal during the CD was not improved in this more realistic case. We found that the non-Gaussianity of the signal was decreased by the inclusion of RSDs during the CD. The bispectrum may contain detectable non-Gaussian information. However, while SKA1-Low is likely to have the sensitivity to detect the bispectrum \citep[e.g.][]{Yoshiura2015SensitivityReionization}, \citet{Watkinson2020} suggested foreground removal will be exceedingly complicated. We leave an analysis of the bispectrum of the 21-cm from the CD (as has already been done for the EoR in \citet{Majumdar2020}) of our simulations for future work.

If an inhomogeneous Ly-$\alpha$ background is present, we found that the inclusion of RSDs dramatically increased both fluctuations in the power spectra and visible features in the images during the first stages of the CD. This boost occurred because the inhomogeneous Ly-$\alpha$ background resulted in the absorption signal being detected from the denser regions, while less signal came from the under-dense regions as they are the last to be decoupled from the CMB. Fluctuations were boosted by the RSDs, like the the Kaiser effect, they also caused over-dense regions to appear denser and emit a signal with a greater magnitude. We concluded that it may be possible to not only image the fluctuations from X-ray heating from the CD but also from Ly-$\alpha$ fluctuations in some scenarios, although this will depend on foreground removal and the first sources to form in the CD. A detailed investigation into the possible fluctuations from the Ly-$\alpha$ background, for example the impact of minihaloes, would be extremely interesting.

The main limitation of this work is that our simulations did not cover the large parameter space associated with the CD. There are a multitude of other possibilities for the properties of the sources X-ray heating and the Ly-$\alpha$ background, although a large amount of is has been eliminated by X-Ray background observations \citep[e.g.][]{Hickox2007,Fialkov2017}, and we previously ruled out QSOs less numerous than our current model having a significant impact. Due to the expense of the simulations analysed here, it was not feasible to explore all possible models; this could only be achieved using a faster method with our simulations as validation. In addition to this, there are many other aspects of the early Universe we do not understand. For example, the metallicity of the IGM, the impact of galaxy evolution, and the nature of very early sources (such as Pop.~III stars). Finally, \citet{Jensen2016TheMeasurements} showed that the anisotropy introduced by RSDs may complicate foreground avoidance techniques for the EoR. The impact of both foreground removal and avoidance also needs to be investigated for this time in order to verify whether the CD with be observable with SKA1-Low. 

Despite these limitations, the inclusion of LoS effects in the results from our simulations has  produce more realistic predictions, and has shown that both statistical detection and the possibility of imaging with SKA1-Low is even more likely than previously thought. The power spectra detection remain distinct for the different models and are boosted on scales where the noise from SKA1-Low was making detectability questionable. Non-Gaussianity remains a good probe of rare X-ray sources, although RSDs do not appear to increase this effect. Finally, the ability of SKA1-Low imaging the first stages of the CD is now a possibility, as the inclusion of RSDs boosts fluctuations from an inhomogeneous Ly-$\alpha$ background.

\section*{Acknowledgements}

We would like to thank Peter Nugent for his useful comments on the paper and Zarija Luki\'c for interesting discussions on this work. This project used resources of the National Energy Research Scientific Computing Center (NERSC), which is supported by the Office of Science of the US Department of Energy under Contract No. DE-AC02-05CH11231. This research was supported by the Exascale Computing Project (17-SC-20-SC), a collaborative effort of the U.S. Department of Energy Office of Science and the National Nuclear Security Administration. This work was also supported by the Science and Technology Facilities Council [grant numbers ST/I000976/1 and ST/P000525/1] and the Southeast Physics Network (SEPNet). GM is supported in part by Swedish Research Council grant 2016-03581. This research was supported in part by the Munich Institute for Astro- and Particle Physics (MIAPP) of the DFG cluster of excellence ``Origin and Structure of the Universe". We acknowledge that the results in this paper have been achieved using the PRACE Research Infrastructure resource Marenostrum based in the Barcelona Supercomputing Center, Spain under Tier-0 project `Multi-scale Reionization'. Some of the numerical computations were done on the Apollo cluster at The University of Sussex. Part of the simulations were performed on resources provided by the Swedish National Infrastructure for Computing (SNIC) at the PDC Center for High Performance Computing in Stockholm. The $N$-body simulation used in this work was completed under the Partnership for Advanced Computing in Europe, PRACE, Tier-0 project PRACE4LOFAR on the TGCC Curie computer. Finally, we would like to thank the referee for helping us improve this work with their constructive comments.

\bibliography{paper}
\end{document}